\documentclass[referee]{aa} 
\usepackage{graphicx}
\usepackage{txfonts}

\usepackage{algorithm}
\usepackage{algpseudocode}
\usepackage{braket}
\usepackage{multirow}
\usepackage{amsmath,amssymb,amsbsy}
\usepackage{eclclass}
\usepackage{fancybox}
\usepackage{longtable}

\newcommand{\upe}{\mathrm{e}}

\newcommand{\uppi}{\mathrm{\pi}}
\newcommand{\kb}{k_\text{B}}
\newcommand{\me}{m_\text{e}}
\newcommand{\p}{{\partial}}

\newcommand{\kB}{k_\text{B}}

\newcommand{\wn}{{\tilde{\nu}}}
\newcommand{\En}{{\tilde{E}}}
\newcommand{\codename}{{\sc Optab}}

\newcommand{\correct}{\color{black}}
\newcommand{\referee}{\color{black}}
\newcommand{\refereee}{\color{black}}

\begin{document}

\title{{\codename}: Public code for generating gas opacity tables for radiation hydrodynamics simulations}
\subtitle{}

\author{
  Shigenobu Hirose\inst{1},
  Peter Hauschildt\inst{2},
  Takashi Minoshima\inst{1},
  Kengo Tomida\inst{3},
  \and
  Takayoshi Sano\inst{4}
}
\titlerunning{{\codename}: Public code for mean opacity tables}
\authorrunning{Hirose et al.}

\institute{
  Japan Agency for Marine-Earth Science and Technology,
  Yokohama, Kanagawa 236-0001, Japan\\
  \email{hirose.shigenobu@gmail.com},
  \and
  Hamburger Sternwarte, Gojenbergsweg 112, 21029 Hamburg, Germany
  \and
  Astronomical Institute, Graduate School of Science, Tohoku University, Aoba,
Sendai 980-8578, Japan
  \and
  Institute of Laser Engineering, Osaka University, Suita, Osaka 565-0871, Japan
}

\date{Received 15/04/2021 ; accepted 03/12/2021}

\abstract{
  We have developed a public code, {\codename},
  that outputs Rosseland, Planck, and two-temperature Planck
  mean gas opacity tables for radiation hydrodynamics simulations in astrophysics.
  The code is developed for modern high-performance computing,
  being written in Fortran 90 and using Message Passing Interface and Hierarchical Data Format, Version 5.
  The purpose of this work is
   to provide a platform on which users can generate opacity tables for their own research purposes.
   Therefore, the code has been designed so that a user can easily modify, change, or add opacity
   sources in addition to those already implemented, which include bremsstrahlung,
   photoionization, Rayleigh scattering, line absorption, and collision-induced absorption.
   In this paper, we provide details of the opacity calculations in our code and
   {\referee present validation tests to evaluate the performance of our code.}

}
    
    \keywords{Opacity -- Radiative transfer -- Methods: numerical}
    
    \maketitle
    %

    \section{Introduction}\label{sec:intro}
    Thanks to recent progress in high-performance computing, radiation hydrodynamics (RHD)
    simulations are now popular in many research fields in astrophysics.
    Accordingly, numerical schemes for RHD simulations have been developed.
    However, it is also crucial to use appropriate opacities in order to
    obtain correct solutions of RHD equations,
    {\correct which is the focus of this study.}
    {\correct To save computational time in RHD simulations, it is fairly common to solve a
      frequency-averaged radiative transfer equation with
    the local thermodynamic equilibrium (LTE) assumption. In this case,}
    Rosseland and Planck mean opacities as functions of temperature
    and pressure are required to perform RHD simulations \citep[e.g.,][]{Hirose:2014}.

    Although the methods for calculating {\correct gas} opacities have been established,
    {\correct significant} effort is required to calculate them in practice.
    For example, one has to gather data pertaining to relevant opacity sources scattered
    throughout the literature or on the web and develop
    complicated interfaces capable of reading {\correct various} types of data.
    Moreover, it is computationally intensive to calculate line opacities from
    very large line lists such as those provided by Exomol\footnote{\texttt{http://www.exomol.com/}}
    \citep{Tennyson:2012}.
    Therefore, researchers {\correct tend to} use publicly available opacity tables.
    In the literature, there are a number of papers that provide opacity tables
    \citep{Cox:1970a,Cox:1970b,Alexander:1975,Alexander:1983,Sharp:1992,Alexander:1994,
      Bell:1994,Pollack:1994,Henning:1996,Helling:2000,Semenov:2003,Ferguson:2005,
      Freedman:2008,Helling:2009,Malygin:2014}, 
    including {\sc Opal} Project\footnote{\texttt{https://opalopacity.llnl.gov}} \citep{Rogers:1992} and
    Opacity Project\footnote{\texttt{http://cdsweb.u-strasbg.fr/OP.htx}} \citep[][]{Seaton:1992}.

    As summarized in Table A.1 in \citet{Semenov:2003}, the various opacity tables
    differ depending on the research purpose and thus 
    differ in composition, opacity sources, ranges of temperature and pressure, and
    outputs.
    Consequently, one cannot always obtain an opacity table that meets
    one's research purpose.
    {\correct One can create a hybrid table that covers larger ranges by combining
      different tables \citep[e.g., Fig. 2 in][]{Hirose:2014}; however,
    it is not guaranteed that the tables will share the same boundary values. }
    As pointed out by \citet{Malygin:2014}, disagreements in opacity values 
    between different tables could be caused not by only different assumptions being applied but also because of the use of
    different sampling points. This makes comparison of different opacity tables difficult
    unless table codes are publicly available. Another issue is that not all public tables are
    updated regularly. 

    For these reasons, we decided to develop a public code, {\correct named {\codename}}, that can be used
     to generate opacity tables
     {\correct using} opacity sources and sampling points chosen by the user.
    In the first version of the code, we consider only gas opacities,
    whose {\correct physics are well understood}. 
    Our code outputs Rosseland, Planck, and two-temperature Planck mean opacities, as well as
    the monochromatic opacity. 
    {\correct The two-temperature Planck mean opacity (Eq. \ref{eq:two-temp-Planck})
      is needed when one considers the case in which the radiation and gas temperatures
      differ, such as an accretion disk irradiated by the central star \citep[e.g.,][]{Hirose:2019}.}
    In the literature, the public code {\sc Dfsynthe} \citep{Castelli:2005} can
    be used to generate opacity tables \citep{Malygin:2014}. Additionally,
    {\sc Opal}\footnote{\texttt{https://www.cita.utoronto.ca/\~{}boothroy/kappa.html}}
    \citep{Rogers:1992} and \citet{Semenov:2003}
    \footnote{\texttt{https://www2.mpia-hd.mpg.de/\~{}semenov/Opacities/opacities.html}} provide
    public codes for generating opacity tables.
    In contrast to these codes,
    our code is {\correct developed} for modern high-performance computing --- it is written in
    Fortran 90 using Message Passing Interface (MPI) and Hierarchical Data Format,
    Version 5 (HDF5)\footnote{\texttt{https://www.hdfgroup.org/solutions/hdf5/}} --- 
    and addresses both readability and portability.
    Furthermore, our code is {\correct available} at
    GitHub\footnote{\texttt{https://github.com/nombac/optab}}
    and allows for continuous updates through contributions by users.
        
    The goal of this code is not to provide complete opacity tables for a
    specific purpose, but to provide a platform on which users can generate opacity
    tables for their own purposes.
    Therefore, while opacity sources initially implemented in our code
    are {\correct basic ones}, the code is written so that
    a user can easily modify, change, or add opacity sources.

    We note that {\codename} does not solve the chemical equilibrium (CE) condition {\correct
    to obtain equilibrium abundances of species that are needed for
    calculating opacities. We assume that
    the user has obtained this solution using an external program.}
    {\correct Recently, public CE solvers for astrophysics, including }
    {\sc FastChem} \citep{Stock:2018} and {\sc Tea} {\correct \citep{Blecic:2016}}, have been made available. 

    In Sect. \ref{sec:code_desc}, the structure of the code and the opacity sources implemented in our code are described. 
    In Sect. \ref{sec:opacity}, the details of the opacity calculations in our code are given.
    In Sect. \ref{sec:reduce_line}, a method to reduce the computational effort in line opacity calculations is described.
    In Sect. \ref{sec:example},    {\referee the results of validation tests to evaluate the performance of our code are presented.}
    Finally, future aspects are discussed in Sect. \ref{sec:future}.

    \section{Code description}\label{sec:code_desc}

    {\codename} is written in Fortran 90 using the MPI library for parallel computing as well as
    the HDF5 library for flexible and efficient {\correct parallel} Input/Output (I/O) of very large atomic or molecular line lists.
    
    In Table \ref{table:source}, we list the opacity sources implemented in the first version
    of {\codename}. All sources are publicly available, and the choice of sources is
    based on {\correct the} {\sc Phoenix} code {\correct \citep{Hauschildt:1999, Allard:1995}}\footnote{\texttt{https://www.physik.uni-hamburg.de/en/hs/group-hauschildt/research/phoenix.html}}.
    Because {\correct each subroutine calculating each opacity is modularized},
    the user can easily modify the code to change or add opacity sources.
    {\correct To calculate opacities, in addition to the opacity sources,
      some supplementary data providing atomic properties are needed,
      which are obtained from the National Institute of Standards and Technology (NIST) Atomic Spectra Database (ASD) levels form\footnote{\texttt{https://physics.nist.gov/PhysRefData/ASD/levels\_form.html}} {\correct \citep{Kramida:2020}}
      and NIST Atomic Weights and Isotopic Compositions with Relative Atomic Masses (AWIC)\footnote{\texttt{https://www.nist.gov/pml/atomic-weights-and-isotopic-compositions-relative-atomic-masses}} \citep{Meija:2016,Berglund:2011}. 
    }

    \begin{table*}
      \caption[]{Opacity sources {\correct implemented in the first version of {\codename}}. Sources of collision-induced absorption are not listed here for brevity.}
      \label{table:source}
      \begin{tabular}{ccccc}
        \hline
        \noalign{\smallskip}
        & Bremsstrahlung & Photoionization & Rayleigh scattering & Line absorption\\
        \hline
        \noalign{\smallskip}
        \begin{tabular}{c}atoms /\\{\color{black}atomic ions}\end{tabular} & \citet{vanHoof:2014} & \begin{tabular}{c}\citet{Verner:1995}\\\citet{Verner:1996}\\{\referee compilation by} \citet{Mathisen:1984}\\{\referee TOPbase}\end{tabular}&& Kurucz     \\
        molecules &  & & &\begin{tabular}{c}HITRAN\\Exomol\end{tabular}    \\
          \hline
          \noalign{\smallskip}
          H          &                   &                &\citet{Lee:2005}&\\
          He         &                   &                &\citet{Rohrmann:2018}&\\
          H$_2$      &                   & \citet{Yan:2001}  &\citet{Tarafdar:1973}&\\
          H$^-$      & \citet{John:1988}    & \citet{Ohmura:1960} &             &\\
          H$_2^-$    & \citet{John:1975}    &                &             &\\
          \noalign{\smallskip}
          \hline
      \end{tabular}
    \end{table*}

    As shown in Table \ref{table:source}, {\codename} currently
    accepts the Kurucz line lists\footnote{\texttt{http://kurucz.harvard.edu/linelists}} for atomic lines,
    and the HITRAN\footnote{\texttt{https://hitran.org/}}
    \citep{Gordon:2017} and Exomol line lists for molecular lines.
    Because these line lists are extremely large and are repeatedly accessed in the code,
    they are converted to HDF5 files {\correct from their original ASCII files} beforehand.
    In addition, HITRAN collision-induced absorption (CIA) data and the supplementary data for atomic
    properties taken from the NIST ASD/AWIC are converted and stored in an HDF5 file, respectively.
    The details of these conversions are described in Appendix \ref{sec:data_prep}.

    Fig. \ref{fig:flowchart} shows the flowchart for {\codename}.
    First, the code reads parameters that determine the opacity sources to be considered, the wave number grid $\wn_k$ [cm$^{-1}$] ($k$ is the grid index), and the line profile evaluation described in Sect. \ref{sec:reduce_line}. Then, it reads CE abundances based on which the opacities are computed, which are prepared by the user for some pairs of temperature and pressure (i.e., ``layers''). The equilibrium abundances needs to be stored in a specified HDF5 format; the {\codename} package provides codes for storing the outputs of {\sc FastChem} and {\sc Tea} in the HDF5 format.

    \begin{figure}
      \centering
      \includegraphics[width=8cm]{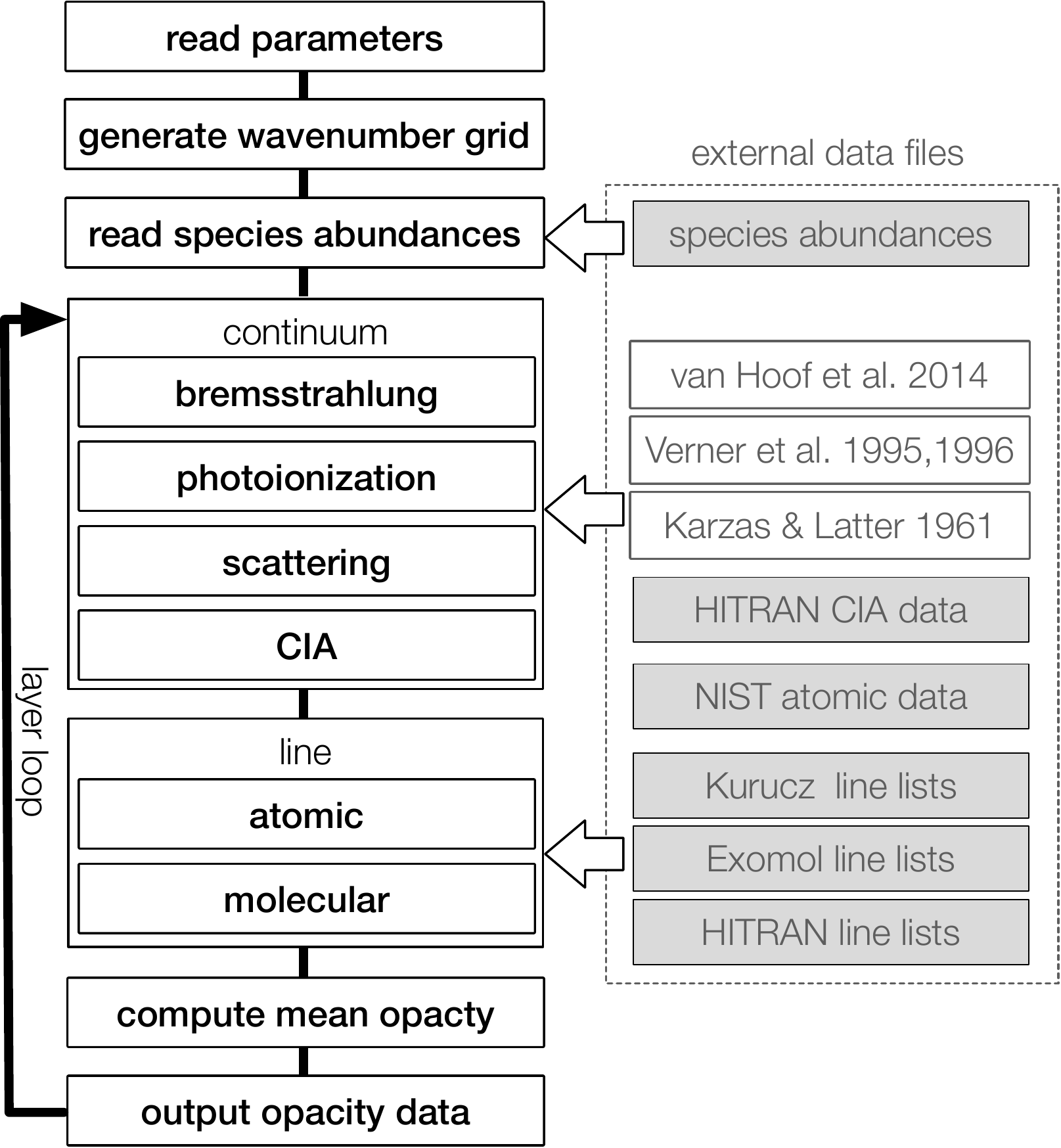}
      \caption{Flowchart of {\codename}.
        The external files in the gray boxes are in the HDF5 format,
        which have been converted from their original ASCII files (Appendix \ref{sec:data_prep}).
      }
      \label{fig:flowchart}
    \end{figure}

    After reading the above data, the code computes the continuum opacities, line opacities,
    and the mean opacities, and {\correct repeats the computations for each layer.
      Here, opacity sources of small size are coded inline, while those of 
    large size are read from external files.}
    MPI parallelization is performed for the layer loop (Sect. \ref{sec:mpi}). 
    As the final step of the layer loop, the absorption and scattering coefficients together
    with the mean opacities {\correct for each layer} are written into a single HDF5 file.


    \section{Opacity calculations}\label{sec:opacity}

    While we use the wavenumber $\wn$ for the computational grid in {\codename},
    we also use the frequency $\nu = c\wn$, photon energy $E = hc\wn$, or
    wavelength $\lambda = 1/\wn$ in the following sections, depending on the context.
    Here, $c$ and $h$ are the speed of light and the Planck constant, respectively. 

    \subsection{Notes on partition functions}
    The opacity calculation for LTE gas uses the fact that the level population obeys
    the Maxwell-Boltzmann distribution, where the number density of a species
    in energy level $l$ at temperature $T$ and pressure $P$ is written as
    \begin{align}
      &N_l(T,P) = N(T,P)\dfrac{g_l\mathrm{e}^{-\frac{E_l - E_1}{\kB T}}}{Q(T)}.\label{eq:Boltz_distr}
    \end{align}
    Here, $Q(T)$ is the internal partition function defined as
    \begin{align}
      Q(T) \equiv \sum_lg_l\mathrm{e}^{-\frac{E_l - E_1}{\kB T}}, \label{eq:pf}
    \end{align}
    $g_l$ and $E_l$ are the statistical weight and energy of the
    level $l$, respectively ({\correct $E_1$ is the energy of the ground state}),
    and $\kB$ is the Boltzmann constant.
    
    The total number density $N(T,P) = \sum_lN_l(T,P)$ in Eq. (\ref{eq:Boltz_distr}) for each species is determined by
    a CE calculation, {\correct where $Q(T)$ also plays
      an important role. This is because} the chemical potential $\mu(T,P)$ [J mol$^{-1}$] used in a
    CE calculation is uniquely related to $Q(T)$ as
    \begin{align}
      \mu(T,P) - N_\text{A}E_1 = RT\ln\left(\dfrac{P}{\kB T}\left(\dfrac{h^2}{2\uppi m\kB T}\right)^\frac32\dfrac{1}{Q(T)}\right), \label{eq:chemical_po}
    \end{align}
    where $N_\text{A}$ is the Avogadro constant [mol$^{-1}$], $R$ is the gas constant [J K$^{-1}$ mol$^{-1}$], and $m$ is the mass of the atom or molecule \citep[Eq. 24.50 in][]{McQuarrie:1997}.
    Therefore, for the LTE opacity to be consistent with the CE, the same
    partition function should be used. For more details, we refer the readers to Appendix \ref{sec:pf}.

    In practice, this cannot be done {\correct for the following} reasons:
    i) In a CE calculation,
    $\mu$ is usually computed from thermochemical databases
    based on experiments such as NIST-JANAF \citep{Chase:1998} or NASA Chemical Equilibrium with Applications (CEA) \citep{McBride:1992}.
    However, because such thermochemical databases do not consider isotoplogues,
    they cannot be used to calculate $Q(T)$ (by means of Eq. \ref{eq:pf_janaf}) for the LTE opacity calculations,
    in which isotopologues are considered. ii) On the other hand, 
    the number of species for which $Q(T)$ is 
     provided along with the line lists is not sufficient for performing a CE calculation.
     {\correct For example, the number of molecular species in the HITRAN line lists is 55},
     whereas that used in a CE solver {\sc FastChem} is 436 \citep[Table 2 in][]{Stock:2018}.

     {\correct Because of reasons i) and ii), we have to admit that in practice
      different partition functions may be used between the CE calculation and the LTE opacity calculations. 
      In our code, for {\referee Kurucz atomic lines}, we use
      {\referee the Kurucz level data \texttt{gf????(z).gam}\footnote{\texttt{http://kurucz.harvard.edu/atoms.html}}}
      to calculate $Q(T)$ from the definition (Eq. \ref{eq:pf}). {\referee We note that the Kurucz level data are not available
      for some species, for which we use the NIST ASD levels instead.}
      For {\referee molecular lines}, we employ
    tabulated partition functions provided with the line lists.

    Hereafter, we omit the dependence of number densities $N$ on temperature $T$ and pressure $P$ for brevity.
    Accordingly, we omit the dependence of absorption coefficients $\alpha$ [cm$^{-1}$] on $P$.

    \subsection{Bremsstrahlung}\label{sec:brems}
    The absorption coefficient of the bremsstrahlung of an atomic ion of charge $Ze$
    (where $Z \ge 1$ is the charge number and $e$ is the elementary charge)
    is written as \citep[e.g.,][]{Rybicki:2008}
    \begin{align}
      \dfrac{\alpha_\text{bre}(\nu;T)}{N_\text{e}N_\text{ion}} &= \left(\dfrac{4}{3}\sqrt{\dfrac{2\uppi \me}{3}}\dfrac{e^6}{\me^2ch}\right)\dfrac{1 - \text{e}^{-\frac{h\nu}{\kb T}}}{\sqrt{\kb T}{\nu}^3}Z^2{\bar{g}_\text{ff}}(\gamma^2,u).\label{eq:brems}
    \end{align}
    {\correct Here, the thermally averaged Gaunt factor, $\bar{g}_\text{ff}$, is a
    function of the scaled quantities $\gamma^2 \equiv {Z^2\text{Ry}}/{\kb T}$
    and $u \equiv {h\nu}/{\kb T}$, and $N_\text{e}$, $N_\text{ion}$, and $m_\text{e}$ are
    the number densities of free electrons and ions, and the electron mass, respectively.
    Ry is an energy unit, the Rydberg, which is approximately $13.6$ eV.}
    Our code employs a numerical table for ${\bar{g}_\text{ff}}(\gamma^2,u)$ provided
    by \citet{vanHoof:2014} to compute Eq. (\ref{eq:brems}) for elements ranging from hydrogen (H) to zinc (Zn).
    {\correct Then, the total coefficient is obtained by summing up the coefficients of
      H$^+$, He$^+$, He$^{2+}$, \dots, and Zn$^{30+}$.}

    In addition, we have implemented the bremsstrahlung absorption of H$^-$ and H$_2^-$
    based on \citet{John:1988} and \citet{John:1978}, respectively.
    In the former, the absorption coefficient per electron pressure ($P_\text{e}$) per number density of H ($N_\text{H}$) [cm$^2$dyne$^{-1}$] is given as a function of $\lambda$ and $T$. In the latter, the absorption coefficient per $P_\text{e}$ per $N_\text{H$_2$}$ is given as a function of $\wn$ and $T$.
    
    \subsection{Photoionization}\label{sec:photoionization}

    A general expression for the photoionization absorption coefficient is 
    \begin{align}
      \alpha_\text{pho}(\nu;T) &= \left(1-\text{e}^{-\frac{h\nu}{\kb T}}\right)\sum_l \sigma_{l} N_{l},
    \end{align}
    which is rewritten using Eq. (\ref{eq:Boltz_distr}) as
    \begin{align}
      \dfrac{\alpha_\text{pho}(\nu;T)}{N}&= \left(1-\text{e}^{-\frac{h\nu}{\kb T}}\right) \dfrac{1}{Q(T)}\sum_l \sigma_{l}{g_{l}\text{e}^{-\frac{E_{l}}{\kb T}}}, \label{eq:photo}
    \end{align}
     where $\sigma_{l}$ is the photoionization cross-section of energy level $l$, {\correct and $(1 - \text{e}^{-{h\nu}/{\kb T}})$ is a correction factor for the induced absorption.}
    
    For atoms and their positive ions, the last factor in Eq. (\ref{eq:photo}) (a summation on the energy levels) is expressed in our code as
    \begin{align}
      &\sum_l \sigma_{l}{g_{l}\text{e}^{-\frac{E_{l}}{\kb T}}} = \sigma_{1}g_1 + \sum_{l\ge2} \sigma_{l}{g_{l}\text{e}^{-\frac{E_{l}}{\kb T}}},\\
      &\sigma_{1} \equiv \begin{cases}
        \sigma^\text{(v96)}(E) & E \le E_\text{max}\\
        \sum_{nm}\sigma_{nm}^\text{(v95)}(E) & E > E_\text{max}
        \end{cases},\label{eq:E_max}
    \end{align}
    where $n$ and {\referee $m$} are the principal quantum number of the shell and the subshell orbital quantum number, respectively.
    
    {\referee Eq. (\ref{eq:E_max}) indicates that} we employ a hybrid of \citet{Verner:1995} and \citet{Verner:1996} ($\sigma_{nl}^\text{(v95)}$ and $\sigma^\text{(v96)}$, respectively) for photoionization originating from the ground level ($l = 1$) {\referee depending on $E_\text{max}$, which} is a parameter given in \citet{Verner:1996} and is specific to the species. \citet{Verner:1995} proposed a fitting formula to the partial photoionization cross-sections based on Hartree-Dirac-Slater calculations for the ground state shells (of elements from H to Zn). \citet{Verner:1996} presented a set of analytic fits to the photoionization cross-sections (of elements from H to Si, and for S, Ar, Ca, and Fe) calculated by the Opacity Project (OP), which are more accurate at low energies. 
    In Fig. \ref{fig:x_MgI}, {\referee the brown curve} shows how the above hybrid photoionization cross-section is applied to Mg atoms, {\correct where $E_\text{max} = 54.90$ eV, or $\tilde\nu = 4.428\times10^5$ cm$^{-1}$.}
    
    \begin{figure}
      \centering
      \includegraphics[width=8cm]{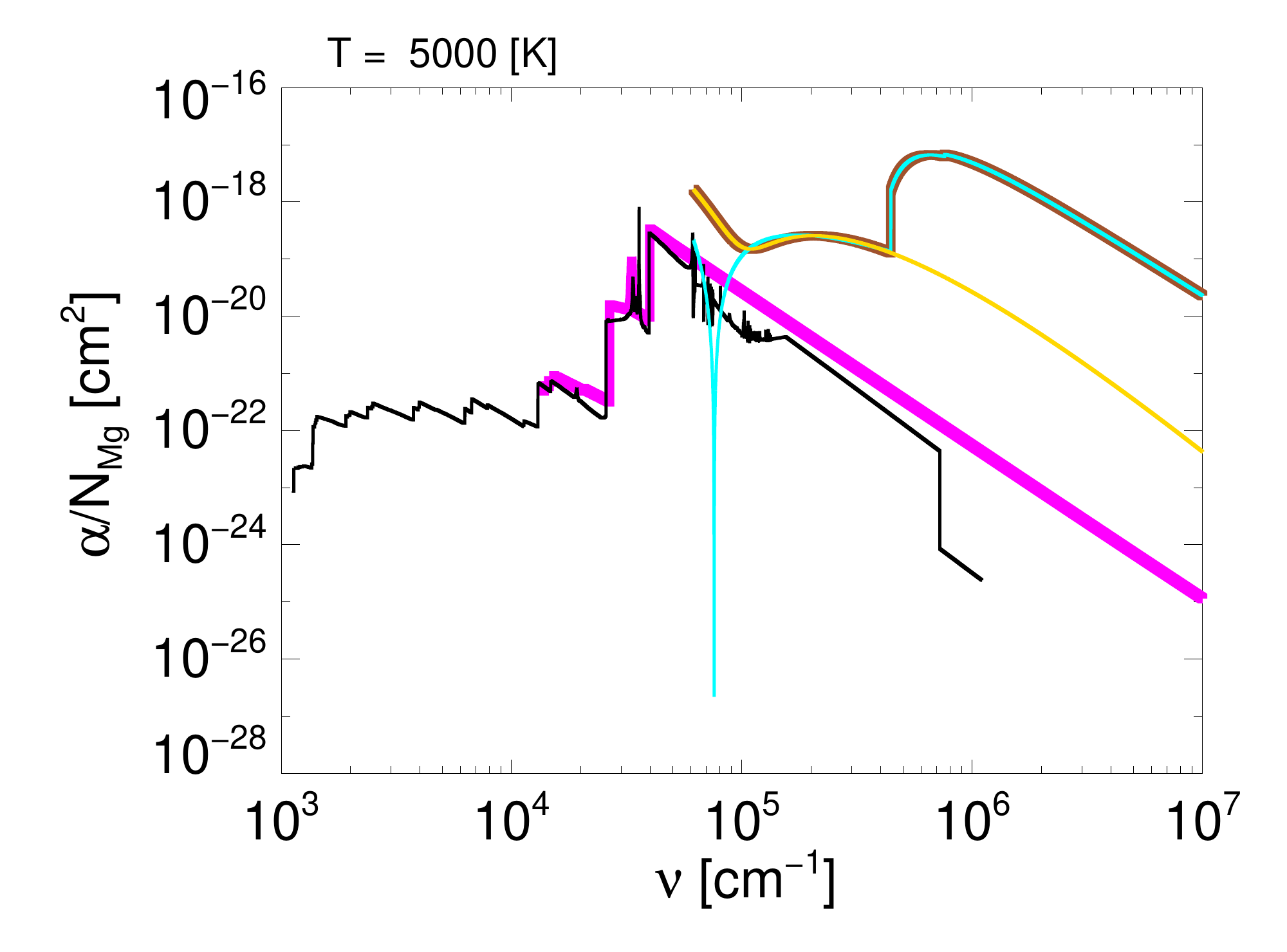}
      \caption{Magnesium photoionization absorption coefficients per Mg atom [cm$^{2}$] at $5000$ K. {\referee The brown curve indicates the photoionization from the ground level, and is a hybrid of \citet{Verner:1996} (outer-subshell contributions; yellow) and \citet{Verner:1995} (inner-subshell contributions; cyan). Both the magenta and black thin curves indicate the photoionization from excited levels; the former is based on the \citet{Mathisen:1984} compilation, whereas the latter is based on TOPbase.}
      }
      \label{fig:x_MgI}
    \end{figure}

    {\referee As for photoionization cross-sections from excited states ($l \ge 2$), {\sc Optab} provides two options: cross-sections compiled by \citet{Mathisen:1984} and cross-sections available at the Opacity Project online atomic database, TOPbase \citep{Cunto:1992}. Both databases provide a set of $\left\{\sigma_{l}, g_{l}, E_{l}\right\}$ for selected photoionization channels of selected species that can be used in Eq. (\ref{eq:photo}).
      TOPbase is generally more comprehensive and more complete than the Mathisen's compilation.\footnote{\refereee We note that some major species, including \ion{Fe}{I} and \ion{Fe}{II}, are not available at the web interface (\texttt{http://cdsweb.u-strasbg.fr/topbase/xsections.html}) but they are available in the ftp version (\texttt{http://cdsweb.u-strasbg.fr/topbase/ftp.html}).} 
      In Fig. \ref{fig:x_MgI}, we compare the photoionization cross-sections from excited states between the Mathisen's compilation \citep[magenta;][]{Vernazza:1976,Thompson:1974,Bradley:1976,Vernazza:1981} and TOPbase (thin black). We observe that more detailed structures are resolved in the TOPbase photoionization cross-sections; however, they require higher computational costs.}

     For H and He$^+$ (hydrogenic atoms and ions), 
     the photoionization cross-section is written in terms of Kramers' semi-classical formula as \citep[e.g.,][]{Carson:1968}\footnote{There is an exact analytic expression of the cross-sections for the ground level of hydrogenic atoms and ions \citep[Eq. (13.1) in][]{Draine:2011}.}
    \begin{align}
      &\sigma_n(\nu) \equiv \left[\dfrac{64\uppi^4}{3\sqrt{3}}\left(\dfrac{\me e^{10}}{ch^6}\right)\left(\dfrac{Z^4}{n^5}\right)\dfrac{1}{\nu^3}\right] g\left(\dfrac{h\nu}{Z^2\text{Ry}},n\right),
    \end{align}
    where $n$ is the principal quantum number.
    {\referee For the Gaunt factor $g(h\nu/Z^2\text{Ry},n)$, fitting formulae by \citet{Chapman:1969,Chapman:1970} for the tables given in \citet{Karzas:1961} are recommended in
    \citet{Mathisen:1984}. However, we use the original tables in our code instead because we found the fitting formulae diverge in some cases.}

    {\correct In addition to the above, we have implemented photoionization absorption for H$^-$ and H$_2$. For the former  and latter, we employ} the cross-section formulas given in \citet{Ohmura:1960} and \citet{Yan:2001}, respectively.

    \subsection{Line absorption}

    The line absorption coefficient for the transition from state $l$ to state $u$ is expressed as
    \begin{align}
      \alpha_\text{line}(\wn;T) &= {N_l}\left(\dfrac{\uppi e^2}{\me c}\right) f_{lu} \dfrac{\phi_{ul}(\wn)}{c}\left(1 - \mathrm{e}^{-c_2\wn/T}\right),
    \end{align}
    where $c_2\equiv hc/\kB$,  $\phi_{ul}(\wn)$ [cm] is the line profile in terms of the wavenumber $\wn$, and $f_{lu}$ is the oscillator strength {\correct from state $l$ to state $u$.\footnote{Exactly speaking, the line profile $\phi_{ul}(\wn)$ also depends on $T$ and $P$ via broadening parameters, which is omitted for brevity here.} }
    Because we consider LTE, the number density at state $l$ of the species being considered can be written as (cf. Eq. \ref{eq:Boltz_distr})
    \begin{align}
      N_l = x^\text{(iso)}N \dfrac{g_l\upe^{-c_2\tilde{E}_l/T}}{Q(T)},
    \end{align}
    where $N$ is the total number density of the species, and $\tilde{E}_l \equiv E_l / hc$ is the level energy in
    terms of the wavenumber.
    We {\correct have introduced} the isotope or isotopologue fraction $x^\text{(iso)}$ (cf. Appendix \ref{sec:data_prep}) because the line transitions are different between different isotope or isotoplogues.
    Then the absorption coefficient is written as
    \begin{align}
      &\dfrac{\alpha_\text{line}(\wn;T)}{N} = x^\text{(iso)} S_{lu}\phi_{ul}(\wn) {\referee \left(1 - \mathrm{e}^{-{c_2\wn}/{T}}\right)},\label{eq:line_abs}\\
      &S_{lu} \equiv{\left(\dfrac{\uppi e^2}{\me c^2}\right) \left(g_l f_{lu}\right) \dfrac{\mathrm{e}^{-c_2\tilde{E}_l/T}}{Q(T)}},\label{eq:strength}
    \end{align}
    where $S_{lu}$ [cm] is the line intensity, and the factor $\left(g_l f_{lu}\right)$ is called the g-f value.
    
    \subsection{Line profile evaluation}
     In our code, we generally apply the Voigt profile, a convolution of the Gaussian and Lorentzian profiles, for the line profile $\phi_{ul}(\wn)$:
    \begin{align}
      \phi_{ul}(\wn) 
      &= \int_{-\infty}^\infty \left\{\dfrac{1}{\sqrt{\uppi}\sigma_\text{G}}\upe^{-\left(({\wn' - \wn_{ul}})/{\sigma_\text{G}}\right)^2}\right\}\left\{ \dfrac{1}{\uppi} \dfrac{\gamma}{(\wn - \wn')^2 + \gamma^2} \right\}d\wn'\nonumber\\
      &= \dfrac{1}{\sqrt{\uppi}\sigma_\text{G}} K(x,y).\label{eq:line_prof}
    \end{align}
    Here, $K(x,y)$ is the Voigt function defined as
    \begin{align}
      &K(x,y) \equiv \dfrac{y}{\uppi}\int_{-\infty}^\infty \dfrac{\upe^{-t^2}}{(x - t)^2 + y^2} dt,\\
      &x \equiv \dfrac{\wn - \wn_{ul}}{\sigma_\text{G}},\\
      &y \equiv \dfrac{\gamma}{\sigma_\text{G}},
    \end{align}
    {\correct and $\wn_{ul} \equiv \wn_{u} - \wn_{l}$ is the wavenumber of the line transition.}
    
     The Gaussian profile arises from the Doppler broadening owing to the thermal and turbulent motion of the absorbing particle. The Gaussian width $\sigma_\text{G}$ [cm$^{-1}$] is expressed as
    \begin{align}
      &\sigma_\text{G} \equiv \dfrac{\sqrt{2\kB T/m} + v_\text{turb}}{c}\wn_{ul},\label{eq:sigma}
    \end{align}
    where $m$ denotes the mass of the absorber. We have not implemented the turbulent velocity $v_\text{turb}$ in our code, but it can be easily implemented by modifying the source code.
    
    The Lorentzian profile arises from both natural broadening and collisional (pressure) broadening caused by neighboring perturbers. The Lorentz width -- the half-width at half-maximum (HWHM) of the Lorentz profile -- $\gamma$ [cm$^{-1}$] can be written as the sum of the two broadening widths:
    \begin{align}
      \gamma = \gamma_\text{nat} + \gamma_\text{coll}.\label{eq:gamma}
    \end{align}

    {\correct For the natural broadening $\gamma_\text{nat}$, if its width is not provided with the line list,
      we substitute the classical damping constant
    \begin{align}
      \gamma_\text{nat} = \dfrac12\dfrac{1}{2\pi c}\left(\dfrac23\dfrac{e^2\omega_{ul}^2}{mc^3}\right)\label{eq:nat_broad},
    \end{align}
    where $\omega_{ul} = 2\uppi c\wn_{ul}$ is the angular frequency of the line transition.}      
    For the collisional broadening $\gamma_\text{coll}$, the molecular line lists in HITRAN or Exomol
    contain the van der Waals broadening width for most species, while
    the Kurucz atomic line list contains both the quadratic Stark broadening and the
    van der Waals broadening widths for most species.
    When the broadening widths are not available in the Kurucz line list,
    we compute them classically or semi-classically, as described in Appendix \ref{sec:classical_widths}.
    
    In our code, we use direct sampling to evaluate the line profile $\phi_{ul}(\wn)$ at a wavenumber grid point $\wn_k$, which means that 
    \begin{align}
      \left.\phi_{ul}(\wn)\right|_{\wn = \wn_k} = \phi_{ul}(\wn_k). 
    \end{align}
    We note that there is another strategy to evaluate the line profile, which is to average the line profile over a grid bin. A comparison between direct sampling and a binned profile is available in \citet{Yurchenko:2018}.

    To evaluate the Voigt function $K(x,y)$, we employ {\referee an algorithm introduced in Sect. 4.3.4.5 in \citet{Aller:1982}:
      \begin{align}
        K(x,y) \approx \begin{cases}
          \displaystyle{yx^{-2}\sum_{j=0}^6c_jx^{-2j} + \left(1+y^2(1-2x^2)\right)\upe^{-x^2}} & y\le10^{-1}, |x|\ge2.5\\
          \displaystyle{\Re\left[\dfrac{\displaystyle{\sum_{j=0}^6a_j\xi^j}}{\displaystyle{\left(\xi^7 + \sum_{j=0}^6b_j\xi^j\right)}}\right]} & \text{otherwise}
        \end{cases},
        \label{eq:voigt}
      \end{align}
      where $\xi \equiv y - ix$ and the coefficients $a_j$, $b_j$, and $c_j$ are listed in Table \ref{table:voigt}.
      The accuracy of Eq. (\ref{eq:voigt}) for a domain of ($0\le x\le 25, 10^{-10}\le y \le10^{2}$) is shown in Fig. \ref{fig:voigt} in terms of relative errors to the reference of \texttt{wofz} in SciPy.
      (Here, we intended to use the same color scale as in \citet{Schreier:2018} so that the accuracy can be directly compared with that of their methods.)
      The maximum error, which occurs around $(x,y) \approx (4,10^{-1})$, is less than 1\%.
      We also measured the computational time of Eq. (\ref{eq:voigt}), where an array of size $101$ was assigned for the $x$ domain ($0\le x\le 25$) while $y$ was being fixed, which is similar
      to the actual line-profile calculation.
      The result is shown on the right-hand side in Fig. \ref{fig:voigt} as a function of $y$, where the reference is the computational time of a Gaussian function. The computational speed of Eq. (\ref{eq:voigt}) is $\sim 2.5$ times slower than that of the Gaussian function for $y < 10^{-1}$ and is $\sim 4.5$ times slower for $y \geqq 10^{-1}$. Both the accuracy and computational speed of Eq. (\ref{eq:voigt}) are satisfactory for our purpose. 
    }

    \begin{table*}
      \caption[]{Coefficients in the approximated Voigt function (Eq. \ref{eq:voigt})}
      \label{table:voigt}
      \begin{tabular}{cccc}
        $j$ & $a_j$ & $b_j$ & $c_j$\\
        \hline
        0 &122.607931777104326&122.607931773875350&0.5641641\\
        1 &214.382388694706425&352.730625110963558&0.8718681\\
        2 &181.928533092181549&457.334478783897737&1.474395\\
        3 &93.155580458138441&348.703917719495792&-19.57826\\
        4 &30.180142196210589&170.354001821091472&802.4513\\
        5 &5.912626209773153&53.992906912940207&-4850.316\\
        6 &0.564189583562615&10.479857114260399&8031.468\\
        \hline
      \end{tabular}
    \end{table*}
    
    \begin{figure}
      \centering
      \includegraphics[width=8cm]{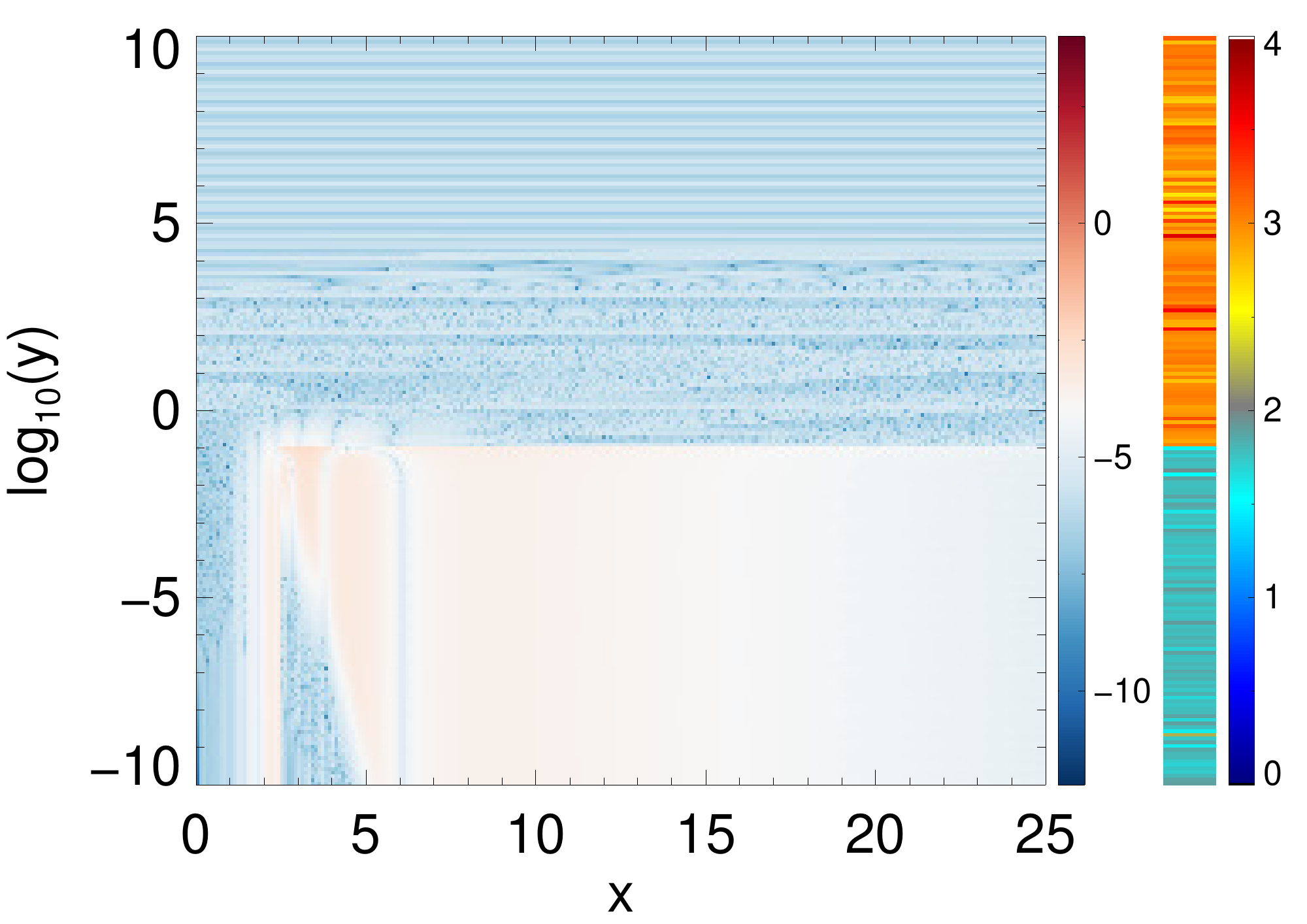}
       \caption{\referee Error of the approximated Voigt function (Eq. \ref{eq:voigt}) relative to the \texttt{wofz} reference (left; logarithmic) and computational time relative to that of the Gaussian function (right; linear). 
      }
      \label{fig:voigt}
    \end{figure}

    \subsection{Collision-induced absorption}
    
    HITRAN provides tabulated coefficients $k_\wn^{(\text{A,B})}$ [cm$^5$ molecules$^{-2}$] for a binary collision complex involving species A and B, from which the total absorption coefficient of CIA is computed as 
    \begin{align}
      {\correct \alpha_\text{CIA}(\nu,T) = \sum_{(A,B)} k_\wn^{(\text{A,B})}N_\text{A}N_\text{B}}.\label{eq:CIA}
    \end{align}
    Here, the coefficients $k_\wn^{(\text{A,B})}$ includes the correction factor of induced absorption.
    In our code, {\correct we consider} pairs of (A,B) = (H$_2$,H$_2$), (H$_2$,He), (N$_2$,N$_2$), (CH$_4$,He), (CO$_2$,CH$_4$), (CO$_2$,CO$_2$), (CO$_2$,H$_2$), (CO$_2$,He), (H$_2$,CH$_4$), (H$_2$,H), (He, H), (N$_2$,H$_2$O),
    (N$_2$,H$_2$), (N$_2$,He), (O$_2$,CO$_2$), (O$_2$,O$_2$), and (O$_2$,N$_2$).

    \subsection{Scattering}\label{sec:scattering}
    With respect to scattering, in addition to Thomson scattering, whose cross-section is
    \begin{align}
      &\sigma_\text{T} = \dfrac{8\uppi}{3}\left(\dfrac{e^2}{\me c^2}\right)^2,
    \end{align}
    we have implemented the Rayleigh scattering of {\correct {H}} in the ground state \citep{Lee:2005}, {\correct H$_2$} \citep{Tarafdar:1973}, and {\correct {He}} in the ground state \citep{Rohrmann:2018}.
    \citet{Lee:2005} reported an exact low-energy expansion of the cross-section redward of Lyman $\alpha$. \citet{Rohrmann:2018} reported analytical expressions for the cross-section including resonances, with reasonable precision up to the 15th resonance. 

    We assume that the scattering is isotropic and coherent, {\correct in which case, any induced scattering is canceled out between the in- and out-scattering.}  Then, the scattering coefficient can be written
    as the sum of the number density $N$ times the scattering cross-section $\sigma^\text{(s)}$: 
    \begin{align}
      {\correct \alpha^\text{(s)}(\wn;T) = N_\text{e}\sigma_\text{T} + \sum_{i = \text{{H},{He}}}{{N_i}_1}\sigma^\text{(s)}_i(\wn) + N_\text{H$_2$}\sigma^\text{(s)}_\text{H$_2$}(\wn),}
    \end{align}
        {\correct where ${N_i}_1 \equiv N_i {g_i}_1/Q_i$ and $\sigma^\text{(s)}_i$ denote the number density in the ground state and the scattering cross-section of species $i$, respectively.}

    \subsection{Mean opacities}\label{sec:mean_opacities}
   {\correct The total absorption coefficient is the sum of all the absorption coefficients described above:}
   \begin{align}
     {\correct \alpha_\text{total} \equiv \alpha_\text{bre} + \alpha_\text{pho} + \alpha_\text{line} + \alpha_\text{CIA}},
   \end{align}
       {\referee where necessary sums related to species and transitions are taken for each term}.

       {\referee The Rosseland mean opacity $\bar\alpha_\text{R}$ and Planck mean opacity $\bar\alpha_\text{P}$ are defined, respectively, as}
    \begin{align}
      &\dfrac{1}{\bar\alpha_\text{R}(T)} \equiv \left. \int\dfrac{1}{\alpha_\text{total}(\wn;T)+\alpha^\text{(s)}(\wn;T)} \dfrac{\p B_\wn(T)}{\p T}d\wn \middle/ \int \dfrac{\p B_\wn(T)}{\p T}d\wn \right.,\\
      &{\correct \bar\alpha_\text{P}(T;T_{2}) \equiv \dfrac{\int \alpha_\text{total}(\wn;T) B_\wn(T_\text{2})d\wn}{\int B_\wn(T_{2})d\wn}},\label{eq:two-temp-Planck}
    \end{align}
    {\referee where $T_2 = T$ (gas temperature) for the normal Planck mean opacity and $T_2 = T_\text{rad}$ (external radiation temperature) for the two-temperature Planck mean opacity}. $B_\wn(T)$ and $\p B_\wn/\p T$ are, respectively, the Planck function in terms of wavenumber $\wn$ and its temperature derivative:
    \begin{align}
      &B_\wn(T) \equiv B_\nu(T)\dfrac{d\nu}{d\wn} = 2hc^2\dfrac{\wn^3}{\exp(c_2\wn/T) - 1},\\
      &\dfrac{\p B_\wn(T)}{\p T} = 2hc^2\dfrac{c_2\wn^4\exp(c_2\wn/T)}{T^2(\exp(c_2\wn/T) - 1)^2},
    \end{align}
    where $B_\nu(T)$ is the Planck function in terms of frequency $\nu$. We note that the dependence {\referee of the opacities} on pressure $P$ is omitted here.

    {\referee
      We use the trapezoid rule to evaluate integrals on wavenumber $\wn$ for the Rosseland mean opacity $\bar\alpha_\text{R}$. As for the Planck mean opacity $\bar\alpha_\text{P}$, the integration of absorption coefficients on the wavenumber $\wn$ strongly depends on the grid resolution, particularly for lines. Therefore, while we evaluate the continuum part of the Planck mean opacity using the trapezoid rule on the wavenumber grid, we evaluate the line part as follows \citep[cf.][]{Mayer:2005}:
      \begin{align}
        &\bar\alpha^\text{line}_\text{P}(T;T_2) \approx \dfrac{\sum_{lu}Nx^\text{(iso)}S_{lu}\left(1 - \exp\left(-c_2\wn_{ul}/T\right)\right)B_{\wn_{ul}}(T_2)}{\int B_\wn(T_{2})d\wn},\label{eq:planck_mean}
      \end{align}
      which is independent of the wavenumber grid. Here, we assume the line profile $\phi_{ul}(\wn) \approx \delta(\wn-\wn_{ul})$, and the summation $\sum_{lu}$ is taken for all the transitions (the summation on the species is omitted for brevity). The validation of this method to evaluate the Planck mean opacity is described in Section \ref{sec:code_parameters}.
    }

    \section{Reducing computational effort in the line opacity calculations}\label{sec:reduce_line}

    The bottleneck in the opacity calculations is the line absorption, particularly the Voigt profile
    calculation. This is because computation of the Voigt function $K(x,y)$ requires significant time
    \citep[e.g., discussion in][]{Schreier:2018}, and because the Voigt profile has wide wings
    and thus the number of grid points over which the function is evaluated is
    relatively large compared to the Gaussian profile.
    {\correct Ideally, the Voigt profile calculation should be performed for all line transitions at all grid points.}
    However,
    the number of transitions in the Kurucz atomic line list is several $10^8$, and that
    in the Exomol database exceeds $10^{10}$ for some species. Clearly, it is not practical to
    evaluate the line profile for all these transitions over grid points that can easily exceed $10^5$.
    In reality, not all transitions are strong enough to be evaluated, and the
    {\correct value of opacity} may be negligible at grid points far from the line center.

    {\correct Following the {\sc Phoenix} code, we have implemented an
      algorithm in our code} to reduce the number of ``Voigt lines'': lines for which the Voigt profile needs to be evaluated.
    The algorithm is represented by the pseudo code in Algorithm \ref{algo:line}.
    Suppose a line transition whose center and
    intensity are $\wn_{ul}$ and $S_{lu}$, respectively. We use the Gaussian profile to evaluate
    the nominal absorption coefficient at the line center as
    \begin{align}
      \alpha_\text{line}^* \equiv \dfrac{x^\text{(iso)}NS_{lu}}{\sqrt{\pi}{\correct \sigma_\text{G}}}.
    \end{align}
    Then we consider the following ratio:
    \begin{align}
      &\delta \equiv \dfrac{\alpha_\text{line}^*}{\alpha_\text{cont}^*},\label{eq:delta}\\
      &{\correct \alpha_\text{cont}^* \equiv \alpha_\text{bre}(\wn_{k^*}) + \alpha_\text{pho}(\wn_{k^*}) + \alpha_\text{CIA}(\wn_{k^*})}.
    \end{align}
    Here $\alpha_\text{cont}^*$ is the continuum absorption coefficient at the grid point ($\wn_{k^*}$) nearest to the line center ($\wn_{ul}$).
    If $\delta < \delta_\text{crit}$ (typically $10^{-4}$ in our code), we consider the line to be too weak
    to contribute the total opacity and discard it.
    If $\delta > \delta_\text{voigt}$ (typically unity in our code), we consider the line to be sufficiently strong 
    and evaluate the Voigt profile at grid points
    in the range {\referee $\wn_{ul} - \Delta_\text{V} \le \wn \le \wn_{ul} + \Delta_\text{V}$,
    where $\Delta_\text{V}$ is the search window (typically $\Delta_\text{V} = 100$} cm$^{-1}$ in our code).
    In other cases, {\correct where $\delta_\text{crit} < \delta < \delta_\text{voigt}$,}
    we consider that only the vicinity of the line center contributes to the total
    opacity and evaluate the Gaussian profile instead of the Voigt profile
    (we call this a ``Gauss line'').
    The search window for Gauss lines (typically $\Delta_\text{G} = 3{\correct \sigma_\text{G}}$ in our code) is usually much smaller than that for Voigt lines.
    {\correct These values of search windows for Voigt lines and Gauss lines, as well as the values of
      $\delta_\text{voigt}$ and $\delta_\text{crit}$, are user-defined parameters.}

    Fig. \ref{fig:nlines} demonstrates how the above
    algorithm considerably reduces the number of the time-consuming Voigt lines
    without introducing notable errors into the total opacity and mean opacities.
    The upper panel shows the dependence on the parameter $\delta_\text{crit}$
    with $\delta_\text{voigt} = 0$ fixed; {\correct that is, all considered lines are Voigt lines here}.
    (For comparison,
    the case in which the HITRAN's dynamic intensity cutoff \citep{Rothman:2013} is used
    instead of $\delta_\text{crit}$ is shown in the leftmost.)
     For example, when $\delta_\text{crit} = 10^{-3}$,
    the number of Voigt lines is reduced to {\referee 53} \% of that in the reference
    ($\delta_\text{crit} = \delta_\text{voigt} = 0$); however, 
    the error in the mean opacity {\correct defined as $\sqrt{\left(\left({\bar\alpha - {\bar\alpha}^\text{(r)}}\right)/{{\bar\alpha}^\text{(r)}}\right)^2}$} as well as that in the  monochromatic
    opacity {\correct  defined as $\sqrt{\left.\sum_k \left(\left({\alpha_k - \alpha^\text{(r)}_k}\right)/{\alpha^\text{(r)}_k}\right)^2\middle/\sum_k\right.}$} are negligible {\correct (specifically,
      the largest error is {\referee 0.0020}, occurring in the Rosseland mean opacity)}.
    {\correct Here,} the superscript (r) and the subscript $k$ denote
    the reference and the grid point index, respectively.
    The lower panel shows the dependence on the parameter $\delta_\text{voigt}$
    with $\delta_\text{crit} = 10^{-4}$ fixed.
    For example, when $\delta_\text{voigt} = 1$, the number of Voigt lines is reduced to 
    {\referee $8$}\% of that in the reference, and the errors are sufficiently low {\correct (specifically,
      the largest error of {\referee 0.0042} occurred in the monochromatic opacity)}.

    \begin{figure}
      \centering
      \includegraphics[width=8cm]{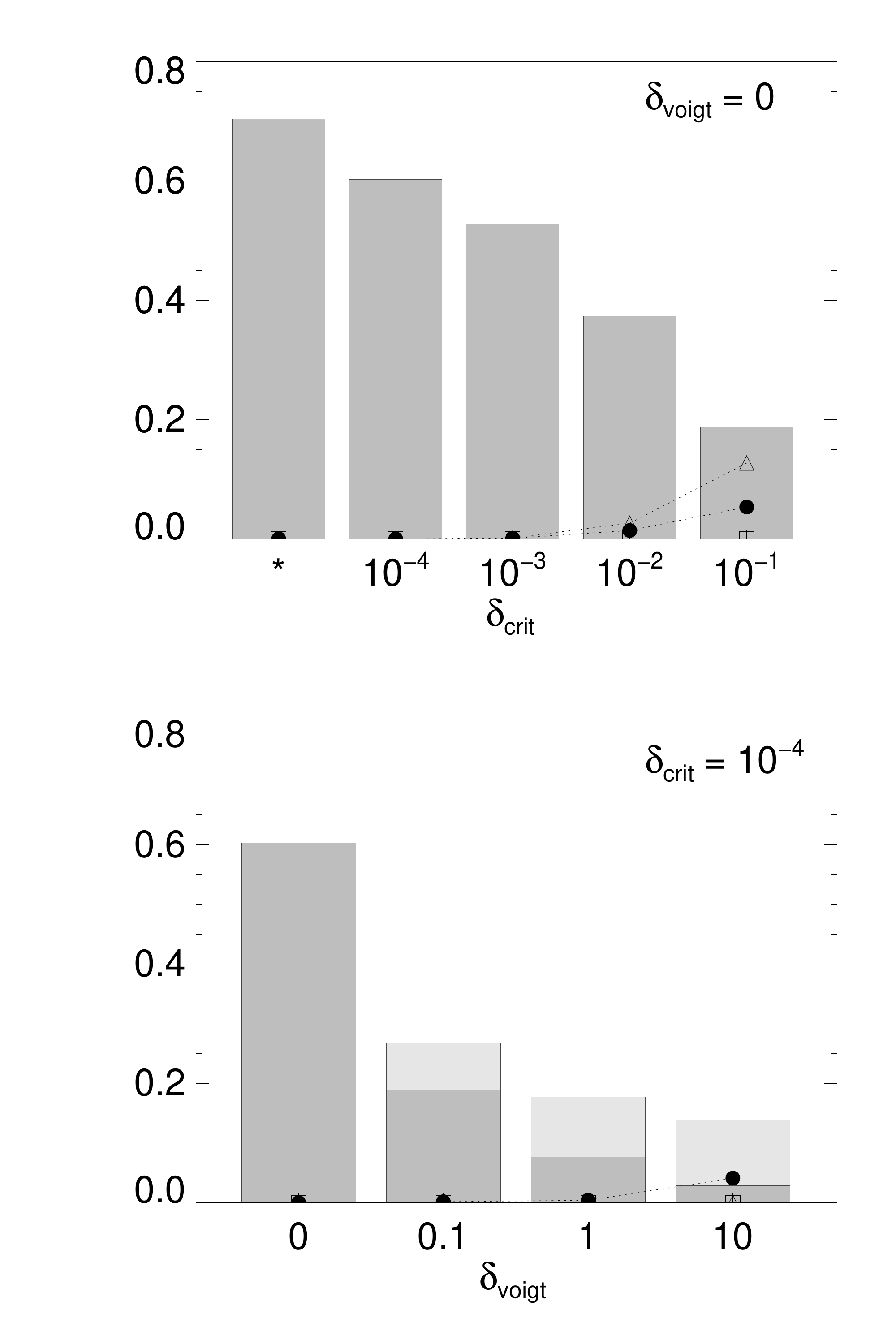}
       \caption{Relative number of the Voigt lines (gray bar) (as well as that of Gauss lines; light gray bar) against the total number of lines in the reference case ($\delta_\text{voigt} = \delta_\text{crit} = 0$). Also, relative errors of the Rosseland (triangle) and Planck (square) mean opacity as well as of the monochromatic absorption coefficient (filled circle) are plotted. {\correct The CE abundance data used here are taken from} the first layer of a demo output produced by {\sc FastChem}, {\tt chem\_output\_agb\_stellar\_wind.dat}.
         {\correct The asterisk in the upper panel indicates a run that employs HITRAN's dynamic strength cutoff, instead of $\delta_\text{crit}$.}
      }
      \label{fig:nlines}
    \end{figure}

    \begin{algorithm}
      \begin{algorithmic}[1] 
        \Procedure{Line}{}
        \State Prepare layers and grid points
        \State Open a line list
        \Repeat
        \State Read a line block from the list
        \ForAll{layers prepared} 
        \If{partition function $Q$ is not defined}
        \State Cycle
        \EndIf
        \If{number density $N = 0$}
        \State Cycle
        \EndIf
        \ForAll{lines in the read block}
        \State Compute line intensity $S$ {\referee{(\ref{eq:strength})}}
        {\referee \State Sum up $S$ for Planck mean (\ref{eq:planck_mean})}
        \State Compute Gaussian width $\sigma_\text{G}$ {\referee{(\ref{eq:sigma})}}
        \State Compute relative line strength $\delta$ {\referee{(\ref{eq:delta})}}
        \If{$\delta < \delta_\text{crit}$}
        \State Cycle
        \ElsIf{$\delta > \delta_\text{voigt}$}
        \State Set profile Voigt
        \State Set search window {\referee $\Delta_\text{V}$}
        \Else
        \State Set profile Gaussian
        \State Set search window {\referee $\Delta_\text{G}$}
        \EndIf
        \State Search grid points in the window
        \If{no grid point in the window}
        \State Cycle
        \EndIf
        \State Compute Lorentz width $\gamma$ {\referee{(\ref{eq:gamma})}}
        \ForAll{grid points in the window}
        \State Compute line profile $\phi$ {\referee (\ref{eq:line_prof})}
        \State Sum up absorption coefficient $\alpha$ {\referee (\ref{eq:line_abs})}
        \EndFor
        \EndFor
        \EndFor
        \Until{end of the line list}
        \State Close the line list
        \EndProcedure
      \end{algorithmic}
      \caption{Computation of the line absorption coefficients. The number in the parentheses indicates the corresponding equation.}
      \label{algo:line}
    \end{algorithm}

    {\referee
    \section{\referee Code validation and performance study}\label{sec:example}

    CE abundances for results presented in this section were obtained using the CE solver {\sc Aces}
    employed in the {\sc Phoenix} code. Here, the element abundance was based on \citet{Asplund:2005} and
    the total number of chemical species considered was 676, which consisted of 428 atomic species (including highly ionized ions) and 248 molecular species.

    \subsection{Comparison with \sc{Phoenix}}
    For the validation of the code, we compared the mean opacities generated by {\sc Optab} with those generated by {\sc Phoenix}. In Fig. \ref{fig:comp_phoenix}, the Rosseland mean, Planck mean, and two-temperature Planck mean (at a radiation temperature of $6000$ K) opacities are compared as functions of temperature for different total pressures. Here, exactly the same wavenumber grid and CE abundances were used. The opacity sources were chosen from those implemented in both {\sc Optab} and {\sc Phoenix}--- specifically, Kurucz atomic lines, Rayleigh scattering by H, He, and H$_2$, and bremsstrahlung and photoionization by atomic ions, including H$^-$. The line evaluation parameters described in Sect. \ref{sec:reduce_line} are set as $\delta_\text{crit} = 10^{-4}$, $\delta_\text{voigt} = 1$, $\Delta_\text{V} = 100$ cm$^{-1}$ ($\Delta_\text{V} = 100$ {\AA} for Phoenix runs), and $\Delta_\text{G} = 3\sigma_\text{G}$. Moreover, here we use the trapezoid rule to evaluate Planck mean opacities (cf. Sect. \ref{sec:mean_opacities}) because {\sc Phoenix} uses that method. 

    Fig. \ref{fig:comp_phoenix} shows that the mean opacities generated by {\sc Optab} basically agree
    with those generated by {\sc Phoenix}. However, there are some discrepancies, particularly in the highest-pressure case ($\log P = 7$), at low temperatures for the Rosseland mean and at high temperatures for the Planck mean. We have identified that the former discrepancy originates from the line profile evaluation, whereas the latter originates from the collisional broadening evaluation.
    The line profile evaluation is performed in the wavenumber basis in {\sc Optab}, whereas it is performed in the wavelength basis in {\sc Phoenix}. As these two bases are identical only near the line center, there may be differences when line profile evaluations are performed for wide ranges. As for the collisional broadening evaluation, when the Kurucz database does not provide the value, {\sc Optab} and {\sc Phoenix} calculate it based on slightly different models (cf. Appendix \ref{sec:classical_widths} for the model employed by {\sc Optab}). The difference is notable when the collisional broadening becomes significant at high pressures.
    When {\sc Optab} employed the same methods as {\sc Phoenix} for evaluating the line profile and the collisional broadening, the generated mean opacities, represented by thin dotted curves in Fig. \ref{fig:comp_phoenix}, agreed with {\sc Phoenix} almost perfectly, as expected.

    \begin{figure}
      \centering
      \includegraphics[width=8cm]{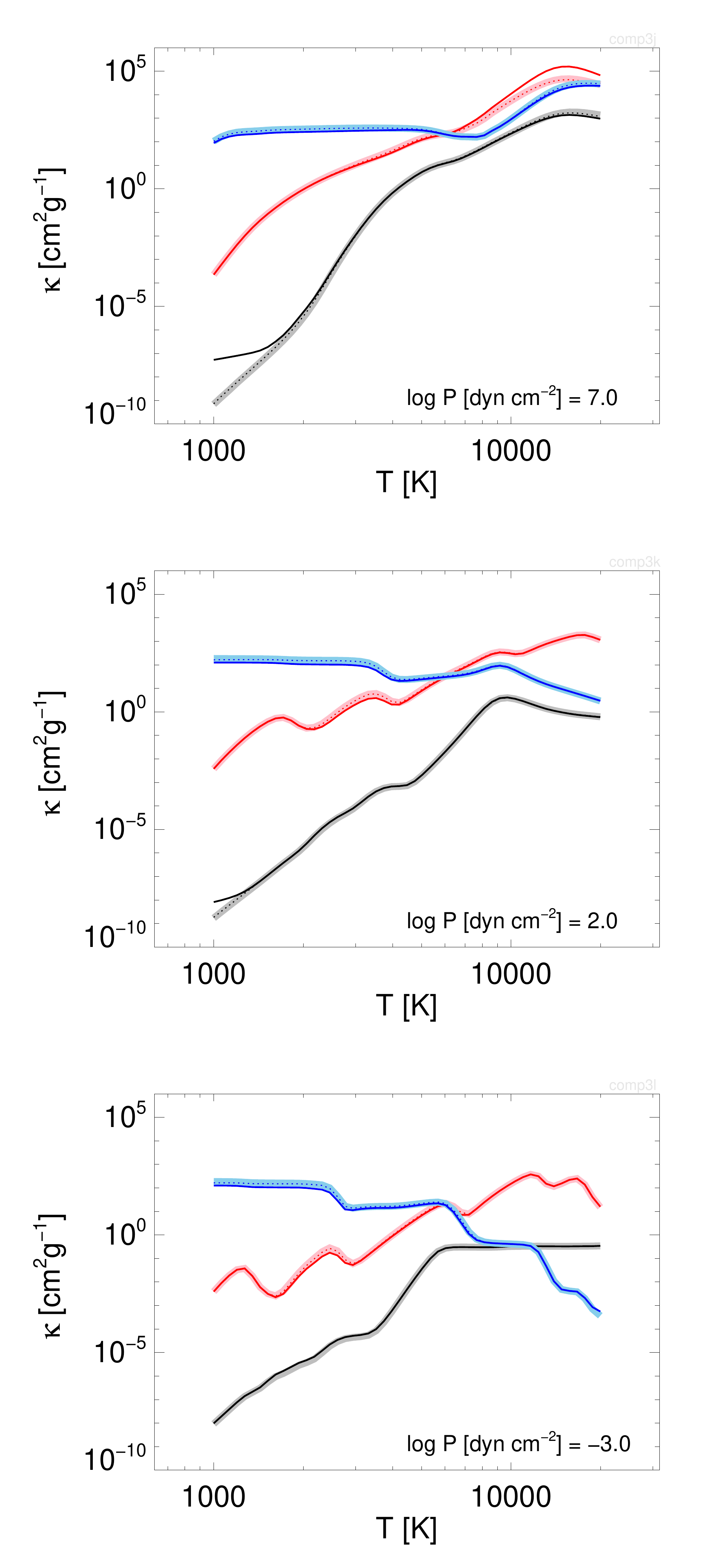}
      \caption{\referee Comparison of mean opacities between {\sc Optab} (thin solid) and {\sc Phoenix} (thick): Rosseland mean (black), Planck mean (red), and two-temperature Planck mean for $T_\text{red} = 6000$K (blue) at $\log P = 7$ (top), $\log P = 2$ (middle), and $\log P = -3$ (bottom). The data represented by the thin dotted curves are explained in the text.
      }
      \label{fig:comp_phoenix}
    \end{figure}
    
    \subsection{Dependence on the number of grid points}\label{sec:code_parameters}

    In this section, we examine the dependence of mean opacities on the grid parameters. Fig. \ref{fig:comp_grid} shows the Rosseland mean, Planck mean, and two-temperature Planck mean opacities at $\log P = 2$ as a function of temperature. The opacity sources we adopted here are Kurucz atomic lines, HITRAN and HITEMP molecular lines, Rayleigh scattering by H, He, and H$_2$, and bremsstrahlung and photoionization by atomic ions, including H$^-$ and H$_2$. The wavenumber grid points were equally spaced on a logarithmic grid covering from $10$ to $10^7$ cm$^{-1}$.

    In each panel, curves of darker colors correspond to calculations using more grid points, ranging from $10^3$ to $10^7$. The Rosseland mean opacities (shown in the top panel) are almost identical when the grid points are $10^4$ or more. This is because the Rosseland mean opacities here are basically determined by the continuum sources. In contrast, the Planck mean and the two-temperature Planck mean opacities (shown in the middle and bottom panels, respectively) strongly depend on the number of the grid points because more line opacities are resolved with more grid points. For the Planck mean opacities (middle panel), as many as $10^6$ grid points are required for convergence, whereas convergence is not observed for the two-temperature Planck mean opacities (bottom panel) with $10^7$ or fewer grid points.

    To avoid the issue of convergence, {\sc Optab} calculates Planck mean and two-temperature Planck mean opacities for lines in a grid-independent way as described in Sect. \ref{sec:mean_opacities}, which are represented by thick light curves in the middle and bottom panels in Fig. \ref{fig:comp_grid}. As evident from the figures, the grid-independent calculation agrees with a grid-dependent calculation with the largest number of grid points for both Planck mean and two-temperature Planck mean opacities, which validates the method of the grid-independent calculation.

    As mentioned in Sect. \ref{sec:reduce_line}, {\sc Optab} has parameters that determines the grid-search window sizes, $\Delta_\text{G} (= 3\sigma_\text{G})$ and $\Delta_\text{V} (= 100 \text{cm}^{-1})$, where the values in the parenthesis are code defaults. We also investigated the dependence of the Rosseland mean opacities on these parameters, which was observed to be weak. As mentioned earlier, this is because the Rosseland mean opacities are mainly determined by the continuum sources in the case investigated here. 

    \begin{figure}
      \centering
      \includegraphics[width=8cm]{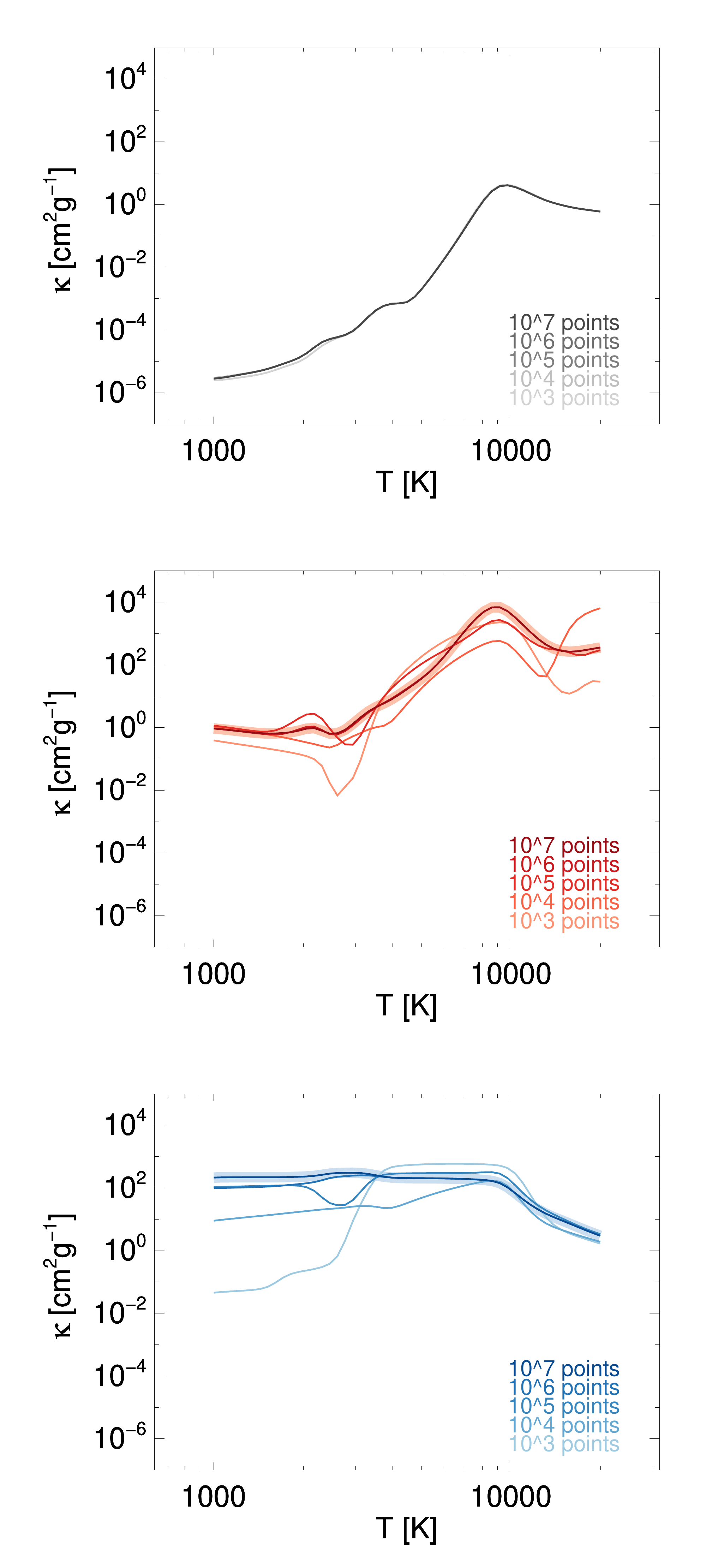}
      \caption{\referee Dependence of Rosseland mean (top), Planck mean (middle), and two-temperature Planck mean for $T_\text{rad} = 6000$K at $\log P = 2$ on the number of grid points, equally spaced on a logarithmic grid. The thick light curves in the middle and the bottom panels are obtained via grid-independent calculations.
      }
      \label{fig:comp_grid}
    \end{figure}

    \subsection{MPI parallelization and its performance}\label{sec:mpi}

    In {\sc Optab}, MPI parallelization on the layer loop (Fig. \ref{fig:flowchart}) has been implemented. The parallelization performance is shown in Fig. \ref{fig:comp_mpi} against the mean opacity calculations for $64$ layers in total (composed of $8$ grid points both in the temperature and pressure grids as shown in Fig. \ref{fig:opactable}). The opacity sources were the same as those described in Sect. \ref{sec:code_parameters}, and the number of the  wavenumber grid points for computing the Rosseland mean opacities was $10^5$. The code parameters used were $\delta_\text{crit} = 10^{-4}$, $\delta_\text{voigt} = 1$, $\Delta_\text{G} = 3\sigma_\text{G}$, and $\Delta_\text{V} = 100$ cm$^{-1}$.

    The dependence of the wall time on the number of MPI processes is shown in the upper panel in Fig. \ref{fig:comp_mpi}. While the computation on each MPI process is independent (i.e., no message passing is necessary), it is often difficult to achieve the load balance between the MPI processes, as shown in the lower panel. This is because the most time-consuming part (i.e., the computation of line opacities) can differ significantly between the layers. Hence, the parallel performance becomes worse as the number of processes increases. We will attempt to improve the parallel performance in the future version of the code. 
    
    \begin{figure}
      \centering
      \includegraphics[width=8cm]{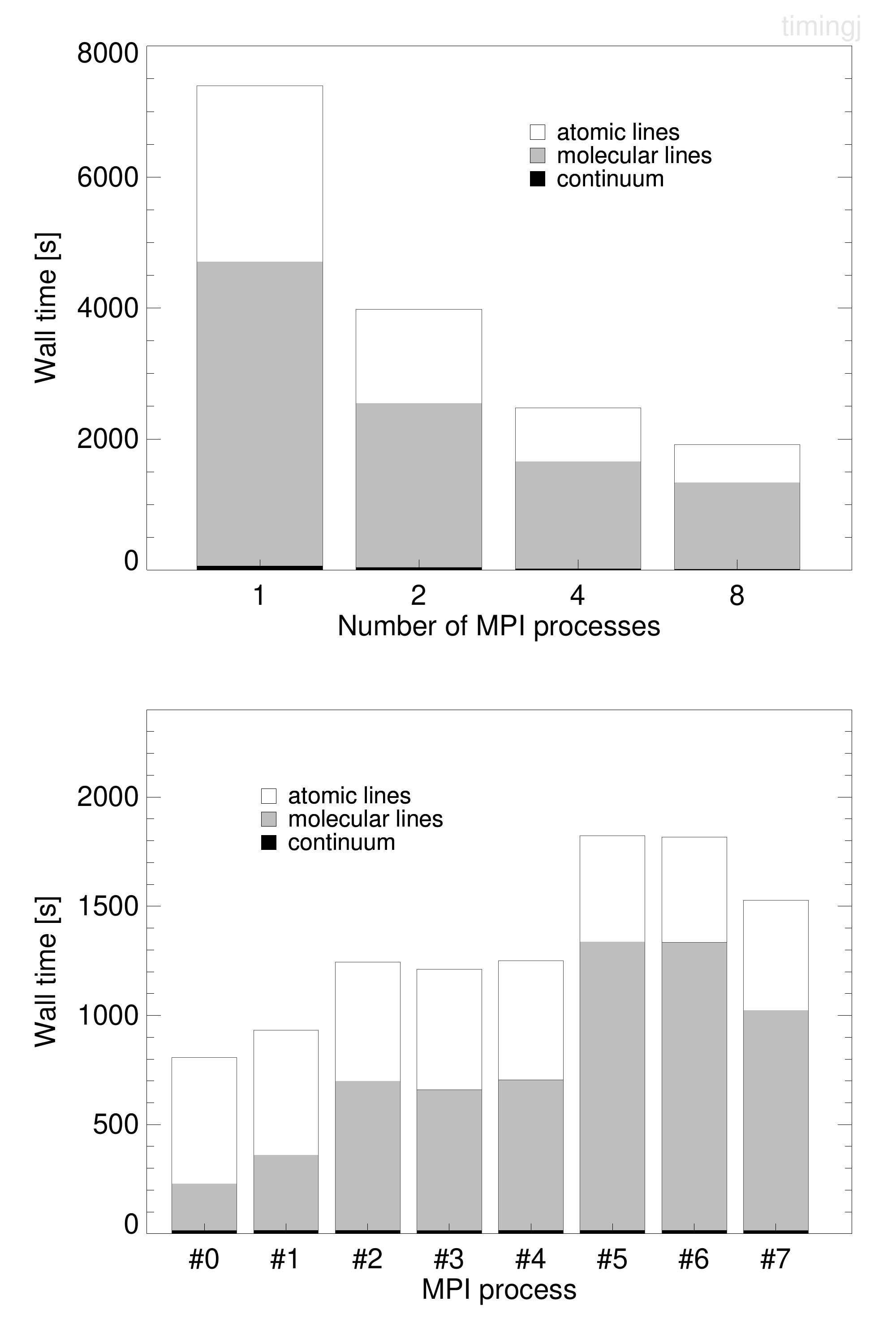}
      \caption{\referee MPI parallel performance for parallelization on the layer loop (upper) and the load balance in the case of 8 MPI processes (lower). The white, gray, and black bars correspond, respectively, to the wall times for atomic lines, molecular lines, and continuum. The computation was performed, using gfortran 10.2 + Open MPI 4.1 + HDF5 1.12, on 2.3 GHz 8-Core Intel Core i9 processors. 
      }
      \label{fig:comp_mpi}
    \end{figure}

    \begin{figure}
      \centering
      \includegraphics[width=8cm]{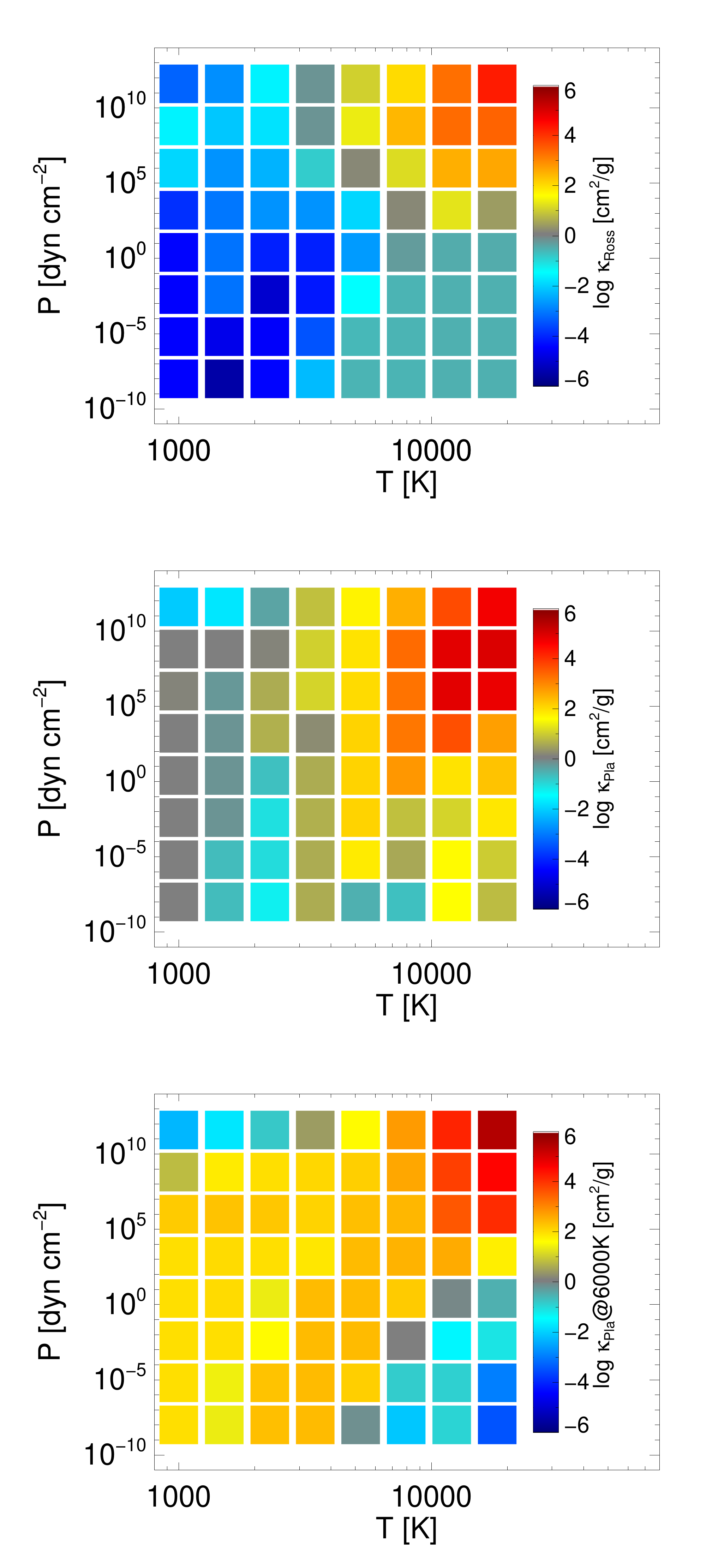}
      \caption{\referee Rosseland mean (top), Planck mean (middle), and two-temperature Planck mean (bottom) opacities on the temperature--pressure grid composed of 64 points (layers).}
      \label{fig:opactable}
    \end{figure}

}

    \section{Discussion on future aspects} \label{sec:future}

    {\correct As demonstrated in the previous section, we have successfully
      developed a code that generates mean opacity tables. 
      The main features are that the code is public and runs 
      based on the user's choice of CE abundances, opacity
      sources, and sampling grid. In this section, we discuss some aspects that remain
      for future work.}
    
    Including dust opacities in our code is obviously the next target.
    We did not include them in the first version of our code because they depend on a dust model, which has not yet been established.
    In addition, the degree to which the dust abundances are involved in the CE of the
    gas mixture is not trivial. 
    (If the dust abundances are independent from the CE of the gas mixture,
    one may simply add dust opacities taken from the literature \citep[e.g.,][]{Semenov:2003}
    to the gas opacities computed by our code, assuming some dust-to-gas mass ratio.)
    We need to solve these issues in order to include dust opacities.

    There are a few issues related to line broadening, although they may not affect the mean opacities.
    For example, we left implementing
    the linear Stark effect for hydrogen atoms \citep[e.g., Sect. 8.2.8 in][]{Hertel:2014} as future work. 
    We are planning to use the tables provided in {\referee \citet{Stehle:1999,Stehle:2010}} to implement the linear Stark effect.
    Another issue is that pressure broadening parameters are not always provided together with a line list. If not provided, we can evaluate them for atoms using the (semi-)classical theory described in Appendix \ref{sec:classical_widths}. However, we are not aware of a proper way to evaluate them for molecules.

    Finally, the opacity sources implemented in the first version of the code are basic ones.
    We will continue to compile additional
    opacity sources to enhance our code, and we also expect to receive user contributions via GitHub.

    \begin{acknowledgements}
      {\refereee We thank the anonymous referee for his/her valuable comments to improve our manuscript as well as our code.}
      This work was supported by JSPS KAKENHI (Grant Number JP18K03716). PHH gratefully acknowledges support from the DFG under grants HA 3457/20-1 and HA
3457/23-1. PHH gratefully acknowledges the support of NVIDIA Corporation with the donation of the Quadro P6000 GPU used in this research.
    \end{acknowledgements}
    
%
%

    \bibliographystyle{aa} 
    \bibliography{export_2021-9-30}
    
    \begin{appendix}
      
      \section{Data conversions of line lists into the HDF5 format}\label{sec:data_prep}

      For flexible and efficient I/O of very large atomic or molecular line lists (Kurucz, HITRAN, and Exomol),
      {\correct their original ASCII files are converted into HDF5 files beforehand.}
      {\correct Furthermore, the atomic properties in ASCII format collected from NIST ASD/AWIC, {\referee TOPbase energy levels and photoionization cross-sections, and}
        HITRAN CIA data are
        converted into a single HDF5 file, respectively.}
      {\correct Programs for these conversions to HDF5
        (written in Fortran 2003) are associated with the code package. }
      Table \ref{table:suppl} summarizes the
      external sources for the isotope or isotopologue fraction,
      partition function, and mass that are needed for making these HDF5 files.
      The structure of the HITRAN CIA file in HDF5 format is simple and thus is not explained here.
      
      \subsection{NIST atomic property file in HDF5 format}
      {\correct Fig. \ref{fig:NIST_HDF5} depicts the data structure in the NIST atomic property file in HDF5 format. The data sets \texttt{nstates}, \texttt{term}, \texttt{ene}, \texttt{gtot}, and \texttt{eneion} are taken from the NIST ASD levels form, whereas \texttt{mass} is taken from NIST AWIC.}
      
      \begin{figure}
        \caption{\correct Structure of the NIST atomic property file in HDF5 format. Data sets with asterisks are arrays of size \texttt{nstates}.}
        \begin{classify}{}
          \class{
            \begin{classify}{\texttt{prop}}
              \class{
                \begin{classify}{\texttt{001.00} (H)}
                  \class{\texttt{nstates} $\cdots$ number of states}
                  \class{\texttt{term}$^*$ $\cdots$ state term}
                  \class{\texttt{ene}$^*$ $\cdots$ $\tilde{E}_i$ [cm$^{-1}$]}
                  \class{\texttt{gtot}$^*$ $\cdots$ $g_i$}
                  \class{\texttt{mass} $\cdots$ $m$ [g]}
                  \class{\texttt{eneion} $\cdots$ $I$ [cm$^{-1}$]}
                \end{classify}
              }
              \class{\texttt{002.00} (He)}
              \class{\texttt{002.01} (He$^+$)}
              \class{ $\cdots$}
              \class{\texttt{092.91} (U$^{91+}$)}
            \end{classify}
          }
        \end{classify}
        \label{fig:NIST_HDF5}
      \end{figure}

      \subsection{Kurucz atomic line lists in HDF5 format}\label{sec:kurucz}
      
      \begin{figure}
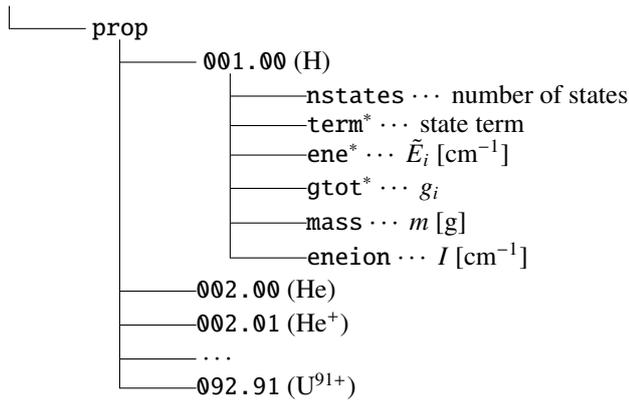

        \caption{\correct Structure of the Kurucz atomic line list in HDF5 format. Data sets with asterisks are arrays of size \texttt{nlines}.}
        \begin{classify}{}
          \class{
            \begin{classify}{\texttt{ trans}}
              \class{\texttt{code}$^*$ $\cdots$ species code}
              \class{\texttt{wnum}$^*$ $\cdots$ $\wn_{ul}$ [cm$^{-1}$]}
              \class{\texttt{lowe}$^*$ $\cdots$ $\tilde{E}_l$ [cm$^{-1}$]}
              \class{\texttt{gfva}$^*$ $\cdots$ $(g_l f_{lu})'$}
              \class{\texttt{gm\_0}$^*$ $\cdots$ $\Gamma^{(\omega)}_\text{natural}\text{[s$^{-1}$]}$}
              \class{\texttt{gm\_1}$^*$ $\cdots$ $\Gamma^{(\omega)}_\text{stark}\text{[s$^{-1}$]}/N_\text{e}\text{[cm$^{-3}$]}$}
              \class{\texttt{gm\_2}$^*$ $\cdots$ $\Gamma^{(\omega)}_\text{vdW}\text{[s$^{-1}$]}/N_\text{H}\text{[cm$^{-3}$]}$}
              \class{\texttt{rsqd}$^*$ $\cdots$ $\braket{R^2/a_0^2}_u$}
              \class{\texttt{nlines} $\cdots$ number of line transitions} 
            \end{classify}
          }
        \end{classify}
        \label{fig:Kurucz_HDF5}
      \end{figure}
      
      In the Kurucz database, there are two categories of line lists for atoms, ``gfall'' and ``gfpred.''
      All lines in the gfall files are based on laboratory-measured energy levels, while the gfpred files contain only predicted lines {\referee to complement the gfall files}.  In Sect. \ref{sec:example}, we used both kinds of the line lists.

      In the Kurucz line lists, each line transition is labeled by its species code\footnote{H = 001.00, He = 002.00, He$^+$ = 002.01, \dots}, and for each line transition, $\tilde{E}_l$, $\tilde{E}_u$, $\log (g_l f_{lu})$ and $\log x_{lu}^\text{(iso)}$, as well as $\Delta\tilde{E}_l$, $\Delta\tilde{E}_u$, and $\Delta\log(g_lf_{lu})$ are provided. The quantities including $\Delta$ are corrections required because of hyperfine shifts. The wavenumber of the line transition is then calculated as
      \begin{align}
        &\wn_{ul} = \left(|\En_u| + \Delta\En_u\right) - \left(|\En_l| + \Delta\En_l\right),
      \end{align}
      where absolute values are used for $\En_u$ and $\En_l$, because in the Kurucz line lists,
      {\correct the level energies have a negative sign when the involving line transition is a predicted (not a measured) one.}
      
      Isotopic shifts in energy levels are miniscule for most atoms, for which the Kurucz line lists do not include distinct isotopes, where $\log x_{lu}^\text{(iso)} = 0$. However, there are a small number of strong lines whose isotopic shifts have been measured in laboratories. In Kurucz's line lists, such lines are indicated by a finite value of $\log x_{lu}^\text{(iso)}$. Therefore, we redefine the g-f values by considering the hyperfine and isotopic shifts as
      \begin{align}
        &(g_lf_{lu})' = 10^{\log(g_lf_{lu}) + \Delta\log(g_lf_{lu})  + \log x_{lu}^\text{(iso)}}\label{eq:gf'}
      \end{align}
      and use this relationship to compute the absorption coefficient (cf. Eq. \ref{eq:line_abs}):
      \begin{align}
        \dfrac{\alpha_\text{line}^\text{Kurucz}({\tilde\nu})}{N} &= S'_{lu}\phi_{ul}(\tilde\nu),\label{eq:line_kurucz}\\
        S'_{lu} &\equiv \left(\dfrac{\uppi e^2}{\me c^2}\right)(g_lf_{lu})'\dfrac{\mathrm{e}^{-c_2\tilde{E}_l/T}}{Q(T)}\left(1 - \mathrm{e}^{-c_2\tilde\nu_{ul}/T}\right).
      \end{align}
      {\correct Because the Kurucz line lists have some incorrect values of $\log x_{lu}^\text{(iso)}$, we replace them in Eq. (\ref{eq:gf'}) with those computed from the NIST AWIC (Table \ref{table:suppl}).}
      
      For line broadening, the Kurucz line lists provide three damping constants (in terms of the angular frequency $\omega$), the natural damping constant $\Gamma^{(\omega)}_\text{nat}$ [s$^{-1}$], the quadratic Stark damping constant (divided by the free electron number density) $\Gamma^{(\omega)}_\text{Stark}/N_\text{e}$, and the damping constant of van der Waals broadening by H atoms (divided by the atomic hydrogen number density) $\Gamma^{(\omega)}_\text{vdW}/N_\text{H}$. 
      The total Lorentzian half-width in terms of the wavenumber is then calculated as
      \begin{align}
        &\gamma = \dfrac12\dfrac{1}{2\uppi c}\left(\Gamma^{(\omega)}_\text{nat} + \left(\dfrac{\Gamma^{(\omega)}_\text{Stark}}{N_\text{e}}\right)N_\text{e} + \left(\dfrac{\Gamma^{(\omega)}_\text{vdW}}{N_\text{H}}\right)N_\text{H}\right).
      \end{align}
      
      {\correct Fig. \ref{fig:Kurucz_HDF5} depicts the data structure in the Kurucz line list in HDF5 format. It shows that the quantities necessary for calculating
        the line opacity Eq. (\ref{eq:line_kurucz}) are stored in the data sets \texttt{wnum},
        \texttt{lowe}, \texttt{gfva}, \texttt{gm\_0}, \texttt{gm\_1}, and \texttt{gm\_2} in the group \texttt{trans}. 
        The data set \texttt{rsqd} is used in Eq. (\ref{eq:rsqd}) in Appendix \ref{sec:classical_widths}.
      }
      The data set \texttt{code} is necessary to identify the species of each line since
      the line transitions of all the species are stored in a single file.
      The partition function for each species is computed from the NIST energy level data, that is, 
        data sets \texttt{ene} and \texttt{gtot} stored in the NIST HDF5 file (Fig. \ref{fig:NIST_HDF5}).
        This is because, in the Kurucz line lists,
        partition functions computed from all measured and predicted levels are provided, but 
        only for a limited number of species.

        \begin{table*}
        \caption{External sources for the isotope or isotopologue fraction, partition function, and mass. }
        \label{table:suppl}
        \begin{tabular}{llll}
          \hline
          & Kurucz & Exomol & HITRAN\\
          \hline
          isotope or isotoplogue fraction & NIST AWIC & NIST AWIC & Isotopologue Metadata\\
          partition function & NIST ASD levels form & --- & Isotopologue Metadata\\
          mass & NIST AWIC & ---  & Isotopologue Metadata\\
          \hline
        \end{tabular}
      \end{table*}

      \subsection{HITRAN and Exomol molecular line lists in HDF5 format}

      Because the isotopic shifts can be significant in molecules, the line list and the partition function table are provided for each isotopologue in databases HITRAN and Exomol. {\correct As for the isotopologue abundances $x^\text{(iso)}$, 
      HITRAN provides them separately from the line list as ``Isotopologue Metadata''\footnote{\texttt{https://hitran.org/docs/iso-meta/}} while Exomol does not provide them. Therefore, for Exomol, we compute $x^\text{(iso)}$ using NIST AWIC (Table \ref{table:suppl}), assuming the terrestrial isotopic abundances of elements}. 
      
      The HITRAN and Exomol line lists give the Einstein coefficients $A_{ul}$ and statistical weights $g_u$, from which the g-f value is calculated as
      \begin{align}
        \left(g_lf_{lu}\right) = \dfrac{c\me}{8\uppi^2e^2\wn_{ul}^2}A_{ul}g_u.
      \end{align}
      
      For line broadening, HITRAN and Exomol consider the van der Waals broadening only and employ the same convention to compute the corresponding Lorentzian half-width as a function of temperature $T$ and the perturber's partial pressure $p$ as
      \begin{align}
        \gamma_\text{vdW}(p,T) = \left(\dfrac{T_\text{ref}}{T}\right)^{n} \gamma_\text{vdW, ref} \left(\dfrac{p}{p_\text{ref}}\right),
      \end{align}
      where $T_\text{ref} = 296$ K and $p_\text{ref} = 1$ ([atm] in HITRAN and [bar] in Exomol) are the reference temperature and pressure, respectively. The exponent $n$ and the reference width $\gamma_\text{vdW, ref} \equiv \gamma_\text{vdW}(p_\text{ref},T_\text{ref})$ [cm$^{-1}$] are parameters given in the line list and depend on the perturbers considered. Both HITRAN and Exomol consider two kinds of perturbers for broadening: HITRAN considers air ($p = p_\text{total} - p_\text{self}$) and self ($p = p_\text{self}$) broadening, while Exomol considers, in most cases, broadening by H$_2$ molecules ($p = p_\text{H$_2$}$) and by He atoms ($p = p_\text{He}$). In our code, we determine the total Lorentzian half-width
      considering both the natural broadening (Eq. \ref{eq:nat_broad}) and the van der Waals broadening by the two kinds of perturbers:
      \begin{align}
        \gamma = \gamma_\text{nat} + \gamma_\text{vdW}^\text{\#1}+ \gamma_\text{vdW}^\text{\#2}.
      \end{align}
      {\correct Here, the perturbers \{\#1, \#2\} denote, respectively, \{``air'', ``self''\} for HITRAN and \{H$_2$, He\} for Exomol.}
      
      \begin{figure}
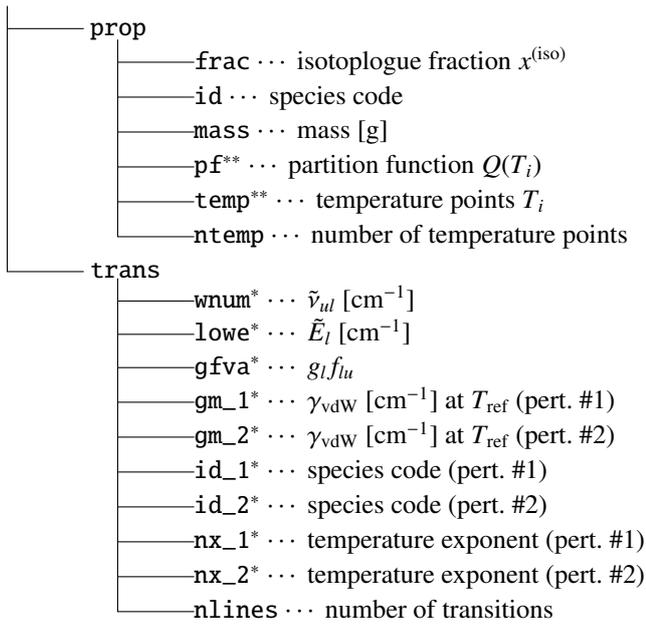

        \caption{\correct Structure of molecular line list in HDF5 format. Data sets with asterisks are
          arrays of size \texttt{nlines}, whereas those with double asterisks are arrays of size \texttt{ntemp}.}
        \label{fig:Exomol_HDF5}
        \begin{classify}{}
          \class{
            \begin{classify}{\texttt{prop}}
              \class{\texttt{frac} $\cdots$ isotoplogue fraction $x^\text{(iso)}$}
              \class{\texttt{id} $\cdots$ species code}
              \class{\texttt{mass} $\cdots$ mass [g]}
              \class{\texttt{pf}$^{**}$ $\cdots$ partition function $Q(T_i)$}
              \class{\texttt{temp}$^{**}$ $\cdots$ temperature points $T_i$}
              \class{\texttt{ntemp} $\cdots$ number of temperature points}
            \end{classify}
          }
          \class{
            \begin{classify}{\texttt{trans}}
              \class{\texttt{wnum}$^*$ $\cdots$ $\wn_{ul}$ [cm$^{-1}$]}
              \class{\texttt{lowe}$^*$ $\cdots$ $\tilde{E}_l$ [cm$^{-1}$]}
              \class{\texttt{gfva}$^*$ $\cdots$ $g_l f_{lu}$}
              \class{\texttt{gm\_1}$^*$ $\cdots$ $\gamma_\text{vdW}$ [cm$^{-1}$] at $T_\text{ref}$ (pert. \#1)}
              \class{\texttt{gm\_2}$^*$ $\cdots$ $\gamma_\text{vdW}$ [cm$^{-1}$] at $T_\text{ref}$ (pert. \#2)}
              \class{\texttt{id\_1}$^*$ $\cdots$ species code (pert. \#1)}
              \class{\texttt{id\_2}$^*$ $\cdots$ species code (pert. \#2)}
              \class{\texttt{nx\_1}$^*$ $\cdots$ temperature exponent (pert. \#1)}
              \class{\texttt{nx\_2}$^*$ $\cdots$ temperature exponent (pert. \#2)}
              \class{\texttt{nlines} $\cdots$ number of transitions} 
            \end{classify}
          }
        \end{classify}
      \end{figure}
      
      {\correct The structure of the molecular line list in HDF5 format for a
        single molecular species is shown in Fig. \ref{fig:Exomol_HDF5}. The species is identified
      by the data set \texttt{id} in the group \texttt{prop}.
      The quantities necessary to compute the line opacity (Eq. {\ref{eq:line_abs}})
      are stored in the data sets in groups \texttt{prop} and \texttt{trans}. The former contains
      quantities specific to the species while the latter contains those specific to the line.
      The partition function is calculated by the linear interpolation of the table provided with the line
      list, which is 
      stored in the data set \texttt{pf} in the group \texttt{prop}. 
      When the temperature to be considered is not within the data set \texttt{temp},
      the temperature range of the partition function, {\correct the line absorption is not considered, as described in line 8 of Algorithm \ref{algo:line}.}
}

      \section{Calculation of collisional damping widths for atomic lines}\label{sec:classical_widths}
      Here we describe how to calculate collisional damping widths for atomic lines
      when they are not given in the Kurucz atomic
      line lists. We consult NIST ASD/AWIC for the ionization energy and mass of atoms,
      which are quantities required to calculate the line broadening parameters in this section.
      
      \subsection{Quadratic Stark broadening}
      According to a semi-empirical formula in \citet[][Eqs. 459, 460]{Griem:1974},
      \begin{align}
        \dfrac{\Gamma_\text{Stark}^{(\omega)}}{2} &= 8\left(\dfrac{\uppi}{3}\right)^\frac32\dfrac{\hbar a_0}{m}N_\text{e}\left(\dfrac{I_\text{H}}{\kB T}\right)^\frac12\nonumber\\
        &\left(\left<\dfrac{R^2}{a_0^2}\right>_u g\left(\dfrac{3\kB T}{2\Delta E_u}\right) + \left<\dfrac{R^2}{a_0^2}\right>_l g\left(\dfrac{3\kB T}{2\Delta E_l}\right)\right),
      \end{align}
      where $I_\text{H}$ is the ionization energy of hydrogen, and  $\Delta E_i$ is the perturbed energy in state $i$ due to a collision. 
      The radiator's effective radius squared (in terms of the Bohr radius $a_0$) in state $i$ is evaluated as
      \begin{align}
        &\left<\dfrac{R^2}{a_0^2}\right>_i = \dfrac12\left(\dfrac{n^*_i}{Z+1}\right)^2\left[5{n^*_i}^2 + 1 - 3l_i(l_i + 1)\right].
      \end{align}
      Here, $l_i$ is the orbital quantum number of the valence electron in state $i$, and the effective principal quantum number in state $i$ is defined as
      \begin{align}
        n^*_i = (Z+1)\left(\dfrac{I_\text{H}}{I - E_i}\right)^\frac12,
      \end{align}
      where $Z$, $I$, and $E_i$ are the charge number, the ionization energy, and the energy in state $i$ of the radiator, respectively.
      
      In our code, for simplicity, we 
      employ the following approximations:
      \begin{align}
        &\left<\dfrac{R^2}{a_0^2}\right>_{\referee i} \approx \dfrac52\dfrac{{n^*_{\referee i}}^4}{(Z+1)^2},\label{eq:rsqd}\\
        &g(x) \approx 0.2.\label{eq:gaunt_approx}
      \end{align}
      {\correct The approximation given by Eq. (\ref{eq:gaunt_approx}) is based on the discussion above pertaining to Eq. 459 in \citet{Griem:1974}.}

      \subsection{van der Waals broadening}
      According to {\correct the Lindholm approximation in classical impact theory \citep[e.g., p.258 in][]{Mihalas:1970},}
      \begin{align}
        \dfrac{\Gamma_\text{vdW}^{(\omega)}}{2} &= N_\text{pert}v_\text{pert}\int_0^\infty \rho\left(1-\cos\left(\dfrac{(3\uppi/8)C_6}{\braket{v}\rho^5}\right)\right)d\rho\nonumber\\
        &= f C_6^\frac25 v_\text{pert}^\frac35 N_\text{pert},\\
        &f \equiv -\left(1+\sqrt{5}\right)\Gamma\left(-\dfrac25\right)\dfrac{\uppi}{5}\left(\dfrac{3\uppi}{8}\right)^\frac25 \approx 8.08,
      \end{align}
      where $N_\text{pert}$ and $v_\text{pert}$ are the number density and the velocity of the perturber, respectively, and $\Gamma(x)$ is the Gamma function.
      Based on quantum theory \citep[e.g., p.297 in][]{Mihalas:1970}, the factor $C_6$ is evaluated as
      \begin{align}
        C_6 = \dfrac{e^2a_0^2}{\hbar}\alpha^{(\text{pol})}_\text{pert} \left(\left<\dfrac{R^2}{a_0^2}\right>_u - \left<\dfrac{R^2}{a_0^2}\right>_l\right),
      \end{align}
      where $\alpha^{(\text{pol})}_\text{pert}$ is the polarizability of the perturber.
      
      In our code, we evaluate the velocity of the perturber as
      \begin{align}
        v_\text{pert} = \sqrt{\braket{v^2}_\text{pert}} = \sqrt{\dfrac{3\kB T}{m^*_\text{pert}}},
      \end{align}
      where $\braket{}$ denotes the thermal average and $m^*_\text{pert}$ is the reduced mass of the perturber with the radiator.
      Although we have implemented van der Waals broadening caused only by atomic hydrogen, {\referee molecular hydrogen, and helium} \citep{Schwerdtfeger:2019,Wilkins:1968}, 
      it is straightforward to add broadening caused by other species if $\left\{v_\text{pert}, N_\text{pert}, \alpha_\text{pert}^{(\text{pol})}\right\}$ are available.
      
      {\correct It is known that }the damping constant evaluated in the above requires some correction. In our code, we multiply $\Gamma_\text{vdW}^{(\omega)}$ by a correction factor of 2.5 based on \citet{Valenti:1996}.

      \section{Partition function, chemical potential, and thermochemical database}\label{sec:pf}

      {\correct In this section,  for reference, we summarize the relationships among the partition function,
        chemical potential, and thermochemical database.}
      
      Chemical equilibrium is obtained by minimizing the Gibbs free energy [J mol$^{-1}$] of the system:
      \begin{align}
        G(T,P) = \sum_i x_i \mu_i(T,P),
      \end{align}
      where $x$ is the molar fraction and index $i$ denotes the species.
      In the chemical equilibrium calculation, the chemical potential $\mu$ [J mol$^{-1}$] is usually computed from
      thermochemical databases as \citep[e.g., Eq.(3.12-5a) in][]{Smith:1982}
      \begin{align}
        \mu(T,P)= RT\ln\left(\dfrac{P}{P^\circ}\right) + RT\left\{\Delta H^\circ_\text{f}(298) - T\left(-\dfrac{{G^\circ}(T) - {H^\circ}(298)}{T}\right)\right\},\label{eq:mu_janaf}
      \end{align}
      where $H$ indicates the total enthalpy and $\Delta H_\text{f}$ indicates the enthalpy of formation,
      and the superscript ${}^\circ$ denotes the standard pressure.
      The quantities $\Delta H^\circ_\text{f}(298)$
      and $-\left({{G^\circ}(T) - {H^\circ}(298)}\right)/{T}$ are provided in thermochemical databases such as NIST-JANAF or NASA CEA.
      We note that {\correct the values of} the chemical potential, Eq. (\ref{eq:mu_janaf}), and Eq. (\ref{eq:chemical_po}) differ by a
      constant because different values of the zero energy are considered.
      
      The partition function $Q(T)$ can be computed using thermochemical databases.
      Hereafter we consider the standard pressure.
      For a pure substance, because the chemical potential is equal to the Gibbs free energy,
      Eq. (\ref{eq:chemical_po}) can be written as \citep[Eq. (24.50) in][]{McQuarrie:1997}
      \begin{align}
        G^\circ(T) - H^\circ(0) = RT\ln\left(\dfrac{P^\circ}{\kB T}\left(\dfrac{h^2}{2\uppi m\kB T}\right)^\frac32\dfrac{1}{Q(T)}\right). \label{eq:mu_q}
      \end{align}
      Then, the partition function is computed as
      \begin{align}
        \ln Q(T)
        &= \left\{-\dfrac32\ln\left(\dfrac{m}{m_\text{u}}\right) - \dfrac52\ln\left(\dfrac{T}{T_1}\right) -\left(\dfrac{S_0}{R} - \dfrac52\right)\right\} \nonumber\\
        &+ {\color{black}\left\{-\dfrac{G^\circ(T)-H^\circ(298)}{T}\right\}}\dfrac{1}{R} + \dfrac{\color{black}\left\{H^\circ(0) - H^\circ(298)\right\}}{RT},\label{eq:pf_janaf}
      \end{align}
      where $m_\text{u}$ is the atomic mass unit, $T_1 = 1$ K, and
      \begin{align}
        \dfrac{S_0}{R} \equiv \ln\left\{\left(\dfrac{2\uppi m_\text{u}\kB T_1}{h^2}\right)^\frac32\dfrac{\kB T_1}{P^\circ}\right\} + \dfrac52 = -1.15170753706
      \end{align}
      is the Sackur–Tetrode constant. The quantity $H^\circ(0) - H^\circ(298)$ is provided in thermochemical databases.

      We note that the partition functions computed
      by Eq. (\ref{eq:pf_janaf}) differ by a factor of the nuclear partition function
      from those provided with the corresponding line list, such as HITRAN or Exomol \citep[][]{Hanson:2015}.
      This is because
      the nuclear partition function is included in the total partition function in the line lists,
      but is not in the thermochemical databases \citep{SousaSilva:2014}.
      We also
      note that the atomic partition functions in our code, which are computed from the NIST ADS levels form,
      do not include the nuclear partition function.

    \end{appendix}
\end{document}